\documentclass{acmtog}

\acmVolume{VV}
\acmNumber{N}
\acmYear{YYYY}
\acmMonth{Month}
\acmArticleNum{XXX}  
\acmdoi{10.1145/XXXXXXX.YYYYYYY}

\acmVolume{28}
\acmNumber{4}
\acmYear{2016}
\acmMonth{October}
\acmArticleNum{106}  
\acmdoi{10.1145/1559755.1559763}

\newcommand{\SCH}{Schr\"{o}dinger }
\newcommand*{\bb}[1]{\boldmath{\mathrm{#1}}}

\usepackage{amsmath}
\usepackage{latexsym}
\usepackage{hyperref}
\usepackage{amsfonts}
\usepackage{color}
\usepackage{comment}
\usepackage[abs]{overpic}
\usepackage{tikz}   
\usepackage{subfigure}
\usepackage{array}
\usepackage{multicol}
\usepackage{multirow}
\usepackage{ulem}

\newtheorem{prop}{Proposition}

\graphicspath{{pics/},{pics/eigenVectors/},{pics/nodalSets/},{pics/texturesDog/},{pics/Kernels/},{pics/spectralCorr/}   }

\begin{document}

\markboth{Y. Choukroun et al.}{Hamiltonian Operator for Spectral Shape Analysis}

\title{Hamiltonian Operator for Spectral Shape Analysis} 

\author{YONI CHOUKROUN, ALON SHTERN, ALEX BRONSTEIN and RON KIMMEL
\affil{Technion - Israel Institute of Technology}
}



\maketitle


\begin{abstract}
Many shape analysis methods treat the geometry of an object as a metric space that can be captured by the Laplace-Beltrami operator. 
In this paper, we propose to adapt the classical Hamiltonian operator from quantum mechanics to the field of shape analysis.
To this end we study the addition of a potential function to the Laplacian as a generator for dual spaces in which shape processing is performed. 
We present a general optimization approach for solving variational problems involving the basis defined by the Hamiltonian using perturbation theory for its eigenvectors.
The suggested operator is shown to produce better functional spaces to operate with, as demonstrated on different shape analysis tasks.
\end{abstract}

\section{Introduction}
The field of shape analysis has been evolving rapidly during the last decades. 
The constant increase in computing power allowed image and shape understanding algorithms to efficiently handle difficult problems that could not have been practically addressed in the past. 
A large set of theoretical tools from metric and differential geometry, and spectral analysis has been imported and translated into action within the shape understanding arena. 
Among the numerous ways of analyzing shapes, a common one is to embed them into a different space where they can be processed more efficiently.

\subsection{Related efforts}
\cite{elad2003bending} introduced a method for analyzing surfaces based on embedding the intrinsic geometry of a given shape into a Euclidean space, extending previous efforts of \cite{schwartz1989numerical,zigelman2002texture,grossmann2002computational}.  
Their key idea was to consider a shape as a metric space, whose metric structure is defined by geodesic distances between pairs of points on the shape. 
Two non-rigid shapes are compared by first having their respective geometric structures mapped into a low-dimensional Euclidean space using \textit{multidimensional scaling} (MDS) \cite{cox2000multidimensional}, and then comparing rigidly the resulting images, also called canonical forms.

\cite{memoli2005theoretical} proposed a metric framework for non-rigid  shape comparison based on the Gromov-Hausdorff distance that was suggested by Gromov as a theoretical tool to quantify disimilarity between metric spaces. 
Using the Gromov-Hausdorff formalism, the distance between two shapes is defined by matching pairwise distances on the shapes. 
However, the Gromov-Hausdorff distance is difficult to compute when treated in a straightforward manner.
To overcome this difficulty \cite{bronstein2006efficient,bronstein2006generalized} proposed an efficient numerical solver based on a continuous optimization problem, known as \textit{Generalized MDS} (GMDS). 
Recently, other relaxation schemes have been proposed, see for example  \cite{chen2015robust,aflalo2013spectral}. 

In the past decade, the \textit{ Laplace-Beltrami operator} (LBO) -- the extension of the Laplacian to non-Euclidean manifolds, has become growingly popular.
Its properties have been well studied in differential geometry and it was used extensively in computer graphics. 
The LBO can be found in countless applications such as mesh filtering \cite{vallet2008spectral}, mesh compression \cite{karni2000spectral}, shape retrieval \cite{bronstein2011shape}, to name just a few.
It has been widely used in shape matching where several approaches treat the correspondence problem by comparing isometric invariant pointwise descriptors between the two shapes. 
For example, the Global Point Signature (GPS) \cite{rustamov2007laplace}, the Heat Kernel Signature (HKS) \cite{sun2009concise} and the Wave Kernel Signature (WKS) \cite{aubry2011wave}, all use the eigenfunctions and eigenvalues of the LBO to compute local shape descriptors. 
Matching only signatures at a small set of points, the correspondence between the points on the two shapes can be found. 
These points can serve as anchors and interpolated for the entire shape \cite{ovsjanikov2010one} where refinement of the basis can be performed to produce precise dense correspondence \cite{ovsjanikov2012functional,pokrass2013sparse,shterniterative}.
 
Recently, learning based approaches \cite{litman2014learning,haoli_CNN,BosMasRodBro16} have also become highly popular in the shape matching arena.

The use of the basis defined by the LBO is in many senses a natural choice for surfaces analysis. 
It was chosen in the functional map framework \cite{ovsjanikov2012functional}  because of its compactness, stability, and invariance to isometries. 
Subsequently, it was proven to be optimal \cite{aflalo2014optimality} for representing smooth functions on the surface. 
In an attempt to overcome the topological sensitivity of the LBO and the non-local support of its eigenfunctions, compressed eigenfunctions have been adapted from mathematical physics to shape analysis \cite{neumann2014compressed,Bron_Chouk_Kim_Sel}.
Here, we try to find a richer family of basis functions that are based on intrinsic properties that can go beyond the geometry of the shape. 
Exploring a similar goal, \cite{kovnatsky2011diffusion} combined geometric and photometric information within a unified metric for shape retrieval.
\cite{iglesias2012schrodinger} used artificial surface textures on shapes to define elliptic operators that give birth to a new family of diffusion distances. Along the same line of thought, \cite{hildebrandt2012modal} designed a new 
 family of eigenvibrations using extrinsic curvatures and deformation energies.

We suggest to further explore those ideas and construct the so-called potential operator that is added to the Laplace Beltrami operator. Here, a designed perturbation to the Laplacian permits a supervised control of the vibrational modes on the manifold.
 
\subsection{Contributions}
 The main contribution of this paper is the exploration of the \textit{Hamiltonian} operator on manifolds.
We study spectral properties of the operator and the impact of an additional potential function to the Laplacian for shape analysis applications.
The properties of the Hamiltonian allow it to be efficiently utilized by many spectral-based methods. 
The potential part can lead to a more descriptive operator when treated as a truncated basis generator. 
Modulated harmonics on the surface are obtained by treating different regions of interest as different values of the potential.
We show that 
by simply plugging the resulting basis into existing spectral shape analysis pipelines
could improve their performance. 

The rest of the paper is organized as follows:
in Section \ref{section:hamiltonian_exploration}, we propose to study 
the Hamiltonian on manifolds from the variational calculus point of view with motivation from quantum mechanics.
We prove optimality properties of its eigenspace, characterize the associated diffusion process, the resulting nodal sets, introduce a discretization method and analyze the robustness of the operator. 

In Section \ref{section:optimization_label}, we propose a global optimization framework for variational problems involving the basis defined by the Hamiltonian. 
We provide an approach for computing derivatives with respect to the potential based on eigenvectors perturbation theory. 
We demonstrate the effectiveness of the framework on the task of data representation.

In Section \ref{section:compressed_modes}, we review recent improvement of the computation of the compressed modes \cite{Ozolins2013} that make use of the decomposition of the Hamiltonian.
Finally, in Section \ref{section:Shape_matching} we present properties of the proposed basis that make it a better alternative for the task of shape matching where priors can be inserted through the potential in order to improve performance.

\section{Hamiltonian operator}
\label{section:hamiltonian_exploration}
\subsection{The Laplace Beltrami Operator}
Consider a parametrized surface $\mathcal{M} : \Omega  \subset \mathbb{R}^2 \to \mathbb{R}^3$ with a metric tensor $(g_{ij})$.
The space of square-integrable functions on $\mathcal{M}$ is denoted by $L^{2}(\mathcal{M})=\{f:\mathcal{M}\rightarrow \mathbb{R} | \int_{\mathcal{M}}f^{2}da<\infty\}$ with the standard inner product $\langle f,g \rangle_{\mathcal{M}} = \int_\mathcal{M} fg \ da$
, where $da$ is the area element induced by the Riemannian metric $\langle\cdot,\cdot\rangle_{g}$.
The Laplace Beltrami Operator acting on a scalar function $f\in L^{2}(\mathcal{M})$ is defined as
\begin{equation}\label{eq:LBO}
\begin{aligned}
\Delta_{\mathcal{M}}f \equiv \mbox{div}_{\mathcal{M}}(\nabla_{\mathcal{M}} f)
  =\frac{1}{\sqrt{g}}\sum_{ij} \partial_{i}(\sqrt{g} g^{ij}\partial_{j}f),
\end{aligned}
\end{equation}
 where $g$ is the determinant of the metric matrix and $(g^{ij}) = (g_{ij})^{-1}$ is the inverse metric. 
If $\mathcal{M}$ is a domain in the Euclidean plane, the metric matrix is generally the identity matrix and the LBO reduces to the well-known Laplacian
\begin{equation}
\begin{aligned}
\Delta f = \frac{\partial^2 f}{\partial x^2}+\frac{\partial^2 f}{\partial y^2}.
\end{aligned}
\end{equation}

  The LBO is self-adjoint and thus admits a spectral decomposition $\{\lambda_{i},\phi_{i}\}$, where $\lambda_{i}\in\mathbb{R}$ and $0=\lambda_{1}\le\lambda_{2}\le ...\uparrow \infty$, such that,
\begin{equation}\label{eq:LBOeigenspace}
\begin{aligned}
 -\Delta_{\mathcal{M}} \phi_i = \lambda_i \phi_i, \\
 \langle \phi_i ,\phi_j \rangle_{\mathcal{M}} = \delta_{ij}.
 \end{aligned}
\end{equation}
with $\delta_{ij}$ the Kronecker delta. 
In case $\mathcal{M}$ has boundary, we add homogeneous Neumann boundary condition
\begin{equation}\label{eq:bound}
\begin{aligned}
\langle \nabla_{\mathcal{M}}\phi_{i},\hat{n}\rangle =0 \ \ \  \text{on} \ \ \ \partial \mathcal{M},
\end{aligned}
\end{equation}
where $\hat{n}$ is the normal vector to the boundary $\partial \mathcal{M}$.

The LBO eigendecomposition can be extracted from the Euler Lagrange solution to Dirichlet energy minimization
\begin{equation}\label{eq:LBOFunctional}
\begin{aligned}
& \underset{\phi_{i}}{\text{min}}
& &\sum_{j=1}^{i} \int_{\mathcal{M}} \|\nabla_{\mathcal{M}}\phi_{j} \|^{2}_{g} \ da, \ \\
& \text{s.t.}
& & \langle \phi_{i},\phi_{j}\rangle_{\mathcal{M}} = \delta_{ij}.
\end{aligned}
\end{equation}
Here, each ordered eigenfunction composing the basis on the manifold corresponds to the function with the smallest possible energy that is orthogonal to all the previous ones.
Therefore, the LBO eigenfunctions can be seen as an extension of the Fourier harmonics in Euclidean spaces to manifolds and are often referred to as \textit{Manifold Harmonics}.

\subsection{Hamiltonian}
\label{section:Hamiltonian}
A Hamiltonian operator $H$ on a manifold $\mathcal{M}$ acting on a scalar function $f\in L^{2}(\mathcal{M})$, is an elliptic operator of the form 
\begin{equation}\label{eq:Hamiltonian}
Hf=-\Delta_{\mathcal{M}}f + Vf,
\end{equation}
where $V:\mathcal{M}\rightarrow \mathbb{R}$ is a real-valued scalar function.
It plays a fundamental role in the field of quantum mechanics appearing in the famous Schr\"{o}dinger equation that describes the wave motion of a particle with mass $m$ under potential $V$,
\begin{equation}\label{eq:SchEquation}
\begin{aligned}
\displaystyle i\hbar\frac{\partial \Psi}{\partial t}  
 = \frac{-\hbar^2}{2m} \Delta  \Psi + V \Psi,
\end{aligned}
\end{equation}
where $\hbar$ is the Planck's constant and $\Psi(x,t)$ represents the wave function of the particle such that $|\Psi(x,t)|^2$ is interpreted as the probability distribution of finding the particle at a given position $x$ at time $t$. \\
The \SCH equation can be analyzed via perturbation theory by solving the spectral decomposition $\{\psi_{i},E_{i}\}_{i=0}^{\infty}$ of the Hamiltonian 
\begin{equation}\label{eq:sch_eigen}
\begin{aligned}
H\psi_{i} = E_i \psi_{i}
\end{aligned}
\end{equation}
also known as the time-independent \SCH equation, 
where $E_{i}$ is the eigenenergy of a particle at 
the stationary eigenstate $\psi_i$.

\begin{figure}[t]
\begin{overpic}[width=1\linewidth,bb=0 0 1280 960,trim={150 50 110 150}]{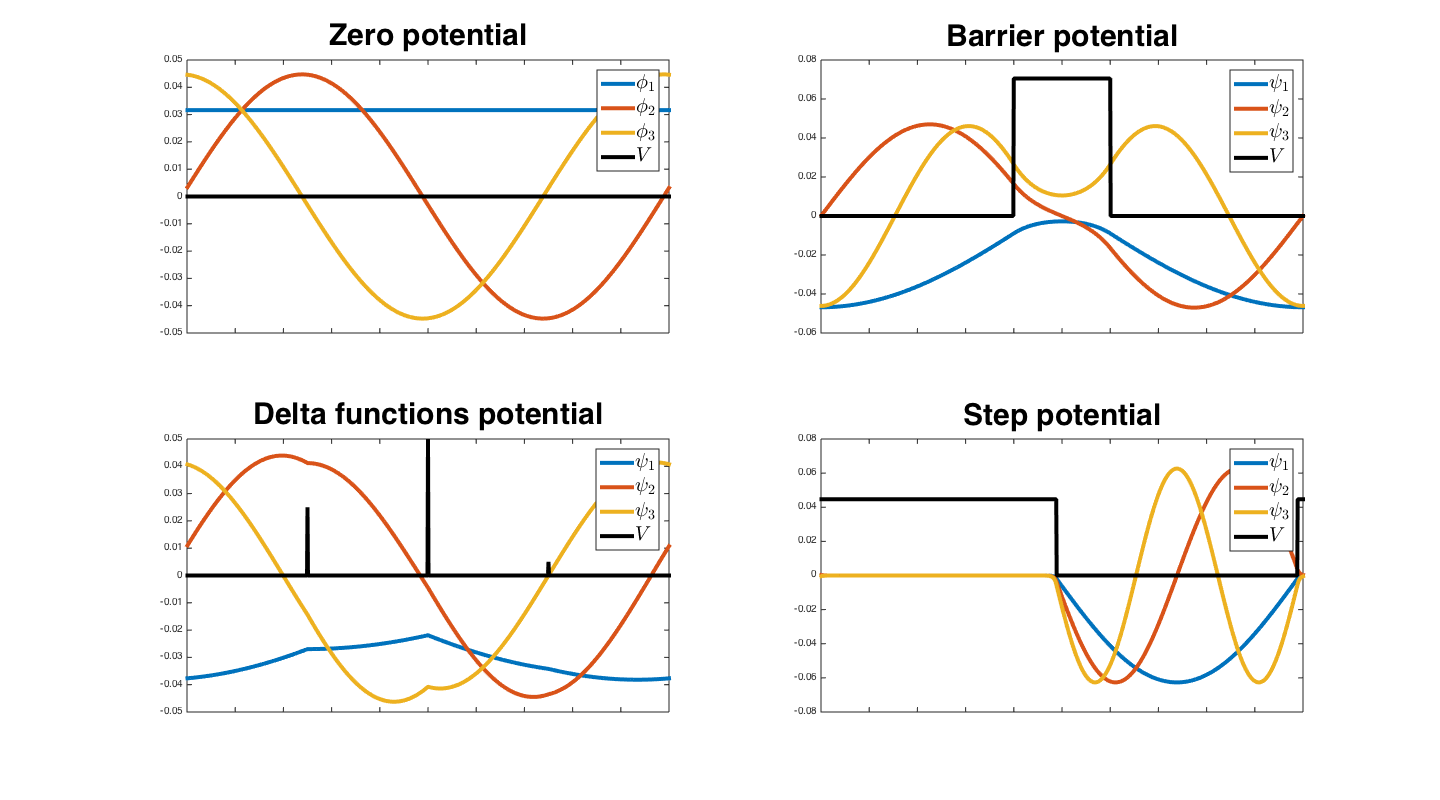}
\centering
 \end{overpic}
 \caption{\label{fig:potentials_euclid} 
Influence of different potentials on the harmonics in one dimension. 
 }
 \end{figure}

Since the potential $V$ is a diagonal operator, the Hamiltonian is self-adjoint as a sum of two self-adjoint operators and its eigenfunctions form a complete orthonormal basis on the manifold $\mathcal{M}$. 
As a generalization of the regular Laplacian, its spectral theory can be derived almost straightforwardly from that of the latter. Classical examples of the influence of potential functions in a one-dimensional Euclidean domain are depicted in Figure \ref{fig:potentials_euclid}.

\subsection{Variational principle}
Let us consider the following variational problem
\begin{equation}\label{eq:SchFunctional}
\begin{aligned}
& \underset{\psi_{i}}{\text{min}}
& &\sum_{j=1}^{i} \int_{\mathcal{M}} 
 \left (\|\nabla_{\mathcal{M}}\psi_{j} \|^{2}_{g} + V \psi_{j}^2\right )  da, \cr
& \text{s.t.}
& & \langle \psi_{i},\psi_{j}\rangle_{\mathcal{M}} = \delta_{ij}, 
\end{aligned}
\end{equation}
whose the Euler-Lagrange equation defines the eigendecomposition of the  Hamiltonian defined in (\ref{eq:sch_eigen}). 

The basis defined by the Hamiltonian operator corresponds to the orthogonal harmonics modulated by the potential function.
The potential defines the trade-off between the orientation and the compactness of the basis and its global support. Larger values of the potential will enforce smooth solutions that concentrate on the low potential regions, while smaller ones will give solutions that better minimize the total energy at the expense of more extended wave functions.


\subsection{Finite step potential}
\begin{figure}[htbp]
\begin{overpic}[width=\linewidth,bb=0 0 1280 960,trim={0 0 0 0},clip]{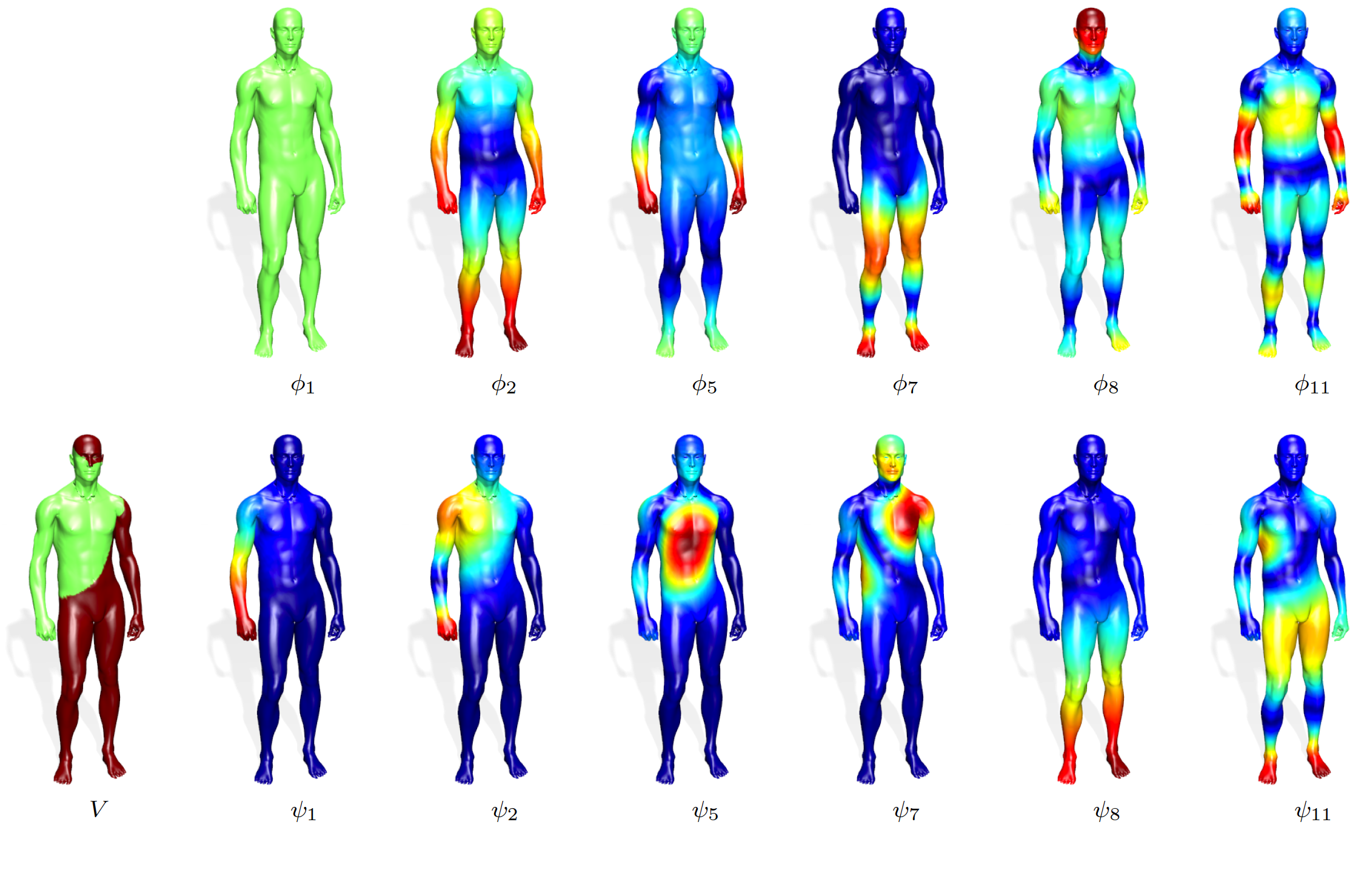}
\end{overpic}
 \centering
\caption{\label{fig:finiteStep} 
Absolute values of the $1^{\mbox{st}},2^{\mbox{nd}},5^{\mbox{th}},7^{\mbox{th}},
 8^{\mbox{th}}$
 and $11^{\mbox{th}}$ eigenfunctions $\{\phi_i\}$ of the LBO (top).
Absolute values of the corresponding eigenfunctions 
 $\{\psi_i\}$ of 
 the Hamiltonian with a step-function potential $V$ (bottom left), with step value $V_{0}$.
For this potential, the first eigenstate $\psi_i$ with energy $E_{i}$  greater than $V_{0}$ is the eighth. 
As analyzed, the eigenfunctions corresponding to lower eigenenergies are restricted to the region with $V=0$, while the higher ones can have effective values (and oscillate) at the $V=V_0>0$ region. An evanescent wave can be observed at the seventh eigenstate. 
}
\end{figure}

The time-independent \SCH equation can yield a rather complicated problem to solve analytically, even in one dimension. 
Let us consider a system with an ideal step potential in one dimension \cite{griffiths2005introduction}. 
We need to solve the differential equation $H\Psi=E\Psi$, with $E$ denoting the energy of the particle, and $V$ the Heaviside function with step of magnitude $V_0>0$, at point $x_0$, given by
\begin{equation}\label{eq:stepPotential}
 \begin{aligned}
 V(x)=
\Bigg\{ \begin{array}{lr}
     \ 0 ,\ \ \ \ \ \ \  x< x_{0}\\
    V_{0} , \ \ \ \ \ \mbox{otherwise}.
  \end{array}
\end{aligned}
\end{equation}
The step divides the space in two constant-potential regions. 
At the zero potential region, the particle is free to move and the harmonic solutions are known. 
In the high potential region, on the other hand, for $E<V_{0}$, the solution is a decaying exponentially, meaning that the particle cannot pass the potential barrier and is reflected according to classical physics. 
If $E>V_{0}$, the solution is also harmonic, which means there is a probability for the particle to penetrate into the effective potential region with a different energy than that of particles in the zero potential region.
We illustrate this effect in Figure \ref{fig:finiteStep} by numerically computing the eigenvectors of the LBO and the Hamiltonian with a potential $V$ defined on a human body surface.

Therefore, the potential energy  can be tuned to enforce localization of the basis at the expense of loss of smoothness.

\noindent 
\begin{theorem} \label{proofEigenvalues}
\textit{
 Let $\{ \phi_{i}, {\lambda}_{i} \}_{i=1}^{\infty}$ and $\{ \psi_{i},E_{i} \}_{i=1}^{\infty}$ be the spectral decompositions of the Laplacian, and the Hamiltonian, respectively. 
Then, $V \geq 0$ everywhere on the manifold implies that the eigenvalues $E_{i}$ satisfy  
\begin{center}
 $ \max_{\mathcal M}(V)+{\lambda}_{i}\geq E_{i}\geq \min_{\mathcal M}(V)+{\lambda}_{i}
 \geq 0$.
\end{center}}
\end{theorem}
\noindent
\textit{Proof.}  
According to the Courant-Fischer min-max theorem, we have
\begin{eqnarray}
E_{i} &=& \underset{\substack{
\Lambda\\
\text{codim} \Lambda = i}}{\text{max}}  \ \      
 \underset{\substack{
 \varphi_{i}\in \Lambda\\
 \varphi_{i} \neq 0}}{\text{min}} 
  \bigg \{ \frac{ \int_{\mathcal M}( \| \nabla_{\mathcal{M}} \varphi_{i} \|^{2}_{g}
   +V \varphi_{i}^{2}) da}{\int_{\mathcal M}\varphi_{i}^{2} da} 
    \bigg \} \cr 
&\geq&  \underset{\substack{
  \Lambda\\
\text{codim} \Lambda = i}}{\text{max}}  \ \ 
 \underset{\substack{ \varphi_{i}\in \Lambda\\
 \varphi_{i} \neq 0}}{\text{min}}
  \bigg\{ \frac{\int_{\mathcal M}(\|\nabla_{\mathcal{M}} \varphi_{i}\|^{2}_{g}+\text{min}_{\mathcal{M}}(V)\varphi_{i}^{2})da}{\int_{\mathcal M}\varphi_{i}^{2} da} \bigg \} \cr
  & & \, \cr
&=&{\lambda}_{i}+\text{min}_{\mathcal M}(V).
\end{eqnarray}
Similarly,
\begin{eqnarray}
E_{i} &\leq&  \underset{\substack{
  \Lambda\\
\text{codim} \Lambda = i}}{\text{max}}  \ \ 
 \underset{\substack{ \varphi_{i}\in \Lambda\\
 \varphi_{i} \neq 0}}{\text{min}}
  \bigg\{ \frac{\int_{\mathcal{M}}(\|\nabla_{\mathcal{M}} \varphi_{i}\|^{2}_{g}+\text{max}_{\mathcal{M}}(V)\varphi_{i}^{2})da}{\int_{\mathcal M}\varphi_{i}^{2} da} \bigg \} \cr
    & & \, \cr
&=&{\lambda}_{i}+\text{max}_{\mathcal M}(V).
\end{eqnarray}
\hfill\(\Box\)\\
Since the family of eigenvalues of the Helmholtz equation (\ref{eq:LBOeigenspace}) consist of a diverging sequence ($\lambda_{n}\propto n \ \text{as} \ n \rightarrow \infty$ \cite{weyl1950ramifications}), there exists an $i$ such that $E_{i}\geq \lambda_{i}+\text{min}_{\mathcal M}(V) \geq \text{max}_{\mathcal M}(V)$ and the trade-off between local-compact and global support of the basis elements can be controlled by the potential energy. Then, we can estimate the magnitude of the potential required in order to allow for oscillations outside the regions where the potential vanishes. 

Given a scalar $\mu \in \mathbb{R}^{+} $ we can define the Hamiltonian as
\begin{equation}
\label{eq:HamiltonianWithBeta}
   H_{\mu}=-\Delta_{\mathcal{M}} +\mu V,
\end{equation}
 where  $\mu$ controls the resistance to diffusion induced  by the potential.
Let $\lambda_{i}$ and $E_{i}$ be the $i$-th eigenvalue of the LBO and Hamiltonian, respectively. We seek for a constant $\mu$ such that 
  $E_{i}>\text{max}_{\mathcal{M}}(\mu V)$ so the particle can penetrate the high potential region.
Considering the potential as small perturbation of the Laplacian, 
 up to first order, the eigenenergies 
 are defined as
 $E_{i} \approx \lambda_{i}+\mu\langle\phi_{i},V\phi_{i}\rangle_{\mathcal{M}}$.
In order to contain the basis support at most until the $i$-th eigenfunction,
 $\mu$ must satisfy
\begin{equation}\label{eq:mu_potential}
\mu < \frac{\lambda_{i}}{\text{max}_{\mathcal{M}}(V)-\langle\psi_{i},V\psi_{i}\rangle_{\mathcal{M}}}.
\end{equation}
According to its potential energy, the basis can then provide supervised multiresolution analysis on the manifold by containing the first eigenfunctions and allow global analysis for the following.
 

\subsection{Optimality of the Hamiltonian eigenspace}

Let us consider a function $f\in L^{2}(\mathcal{M})$.
We define the representation residual function as 
\begin{equation}\label{eq:residual1}
\begin{aligned}
\|r_n\|_{\mathcal{M}}^{2}&=\left\|f-\sum_{i=1}^{\infty}\langle f,{\phi}_{i}\rangle_{\mathcal{M}} {\phi}_{i}\right \|_{\mathcal{M}}^{2} \\ &= \left\|\sum_{i=n+1}^{\infty}\langle f,{\phi}_{i}\rangle_{\mathcal{M}} {\phi}_{i}\right \|_{\mathcal{M}}^{2}=\sum_{i=n+1}^{\infty}\langle f,{\phi}_{i}\rangle^{2}_{\mathcal{M}}.
\end{aligned}
\end{equation}
Defining $\|\nabla_{g}f\|_{\mathcal{M}}^{2}=\int_{\mathcal{M}}\|\nabla_{g}f\|_{g}^{2}da$, we know that
\begin{equation}\label{eq:residual2}
\begin{aligned}
&\|\nabla_{g}f\|_{\mathcal{M}}^{2}+\|\sqrt{V}f\|^{2}_{\mathcal{M}} =\int_{\mathcal{M}}(-\Delta_{\mathcal{M}} f+Vf)f \  da \\ &=\sum_{i=1}^{n}\int_{\mathcal{M}}(\langle f,\psi_{i}\rangle_{\mathcal{M}}E_{i}\psi_{i}) f \ da = \sum_{i=1}^{\infty}E_{i}\langle f,\psi_{i}\rangle_{\mathcal{M}}^{2} \\ &\geq 
\sum_{i=n+1}^{\infty}E_{i}\langle f,\psi_{i}\rangle_{\mathcal{M}}^{2} \geq 
E_{n+1}\sum_{i=n+1}^{\infty}\langle f,\psi_{i}\rangle_{\mathcal{M}}^{2}.
\end{aligned}
\end{equation}
Thus, from (\ref{eq:residual1}) and (\ref{eq:residual2}) we obtain
\begin{equation}\label{eq:represent}
\begin{aligned}
 \|r_n\|_{\mathcal{M}}^{2}=\left \|f-\sum_{i=1}^{n}\langle f,\psi_{i}\rangle_{\mathcal{M}} \psi_{i}\right \|_{\mathcal{M}}^{2}\leq 
  \frac{ \|\nabla_{g}f \|_{\mathcal{M}}^{2}+\|\sqrt{V}f\|_{\mathcal{M}}^{2}}{E_{n+1}}.
\end{aligned}
\end{equation}
Recall that for $V=0$ we return to the LBO case.
Among the numerous reasons that motivated the selection of the Laplacian for shape analysis, a major one is its efficiency in representing functions with bounded gradient magnitude.
This result was subsequently proved to be optimal for representing functions with bounded gradient magnitude over surfaces in \cite{aflalo2014optimality}, which says that there exists no other basis with better representation error for all possible $L^2(\mathcal{M})$ functions.

In case of the Hamiltonian, the Dirichlet energy is coupled with the potential energy. Thus the Hamiltonian operator advocates measuring smoothness differently for different regions of the domain where smoothness remains a less important factor than avoiding vibrations in high potential areas. This is a useful property to exploit in different shape analysis scenarios.

Next, we show that the Hamiltonian is optimal in approximating functions with both bounded gradient and low values in high potential areas.
\begin{theorem} \label{H_Optimality}
\textit{Let $0\leq\alpha<1$. There is no integer $n$ and no sequence $\{\psi_{i}\}_{i=0}^{\infty}$
of linearly independent functions in $L_{2}(\mathcal{M})$ such that}
\begin{eqnarray}
 \left \|f-\sum_{i=1}^{n}\langle f,\psi_{i}\rangle_{\mathcal{M}} \psi_{i}\right \|_{\mathcal{M}}^{2}\leq 
  \frac{\alpha \left ( \|\nabla_{g}f \|_{\mathcal{M}}^{2}+\|\sqrt{V}f\|_{\mathcal{M}}^{2}\right )}{E_{n+1}}   \ \forall f.
\end{eqnarray}
\end{theorem}
The proof of Theorem \ref{H_Optimality} is given in the Appendix.

\subsection{Diffusion process}
Let us be given a Riemannian manifold $\mathcal{M}$. 
The heat equation governing the diffusion process on $\mathcal{M}$ is defined as
\begin{equation}\label{eq:heatKernel}
\Bigg\{ \begin{aligned}
&\partial_{t}u(x,t) = \Delta_{\mathcal{M}}u(x,t), \ \ \ \forall x \in \mathcal{M},\\
&u(x,0) = u_{0}(x),
\end{aligned}
\end{equation}
 with appropriate boundary conditions. 
A natural extension to the new operator with a potential $V$, can be written as
\begin{equation}\label{eq:heatKernelH}
\Bigg\{ \begin{aligned}
&\partial_{t}u(x,t) =Hu(x,t)= \Delta_{\mathcal{M}}u(x,t)-V(x)u(x,t)\\
&u(x,0) = u_{0}(x).
\end{aligned}
\end{equation}
The solutions of (\ref{eq:heatKernel}) and (\ref{eq:heatKernelH}) have the form \cite{iglesias2012schrodinger}
\begin{equation}
\label{eq:heatKernelSol}
 u(x,t) = \int_{\mathcal{M}}u_{0}(y)K(x,y,t)da(y),
\end{equation} that represents the diffusion in time of heat on the manifold $\mathcal{M}$ with potential $V$, where $K(x,y,t)=\sum_{i}e^{-{E}_{i}t}\psi_{i}(x)\psi_{i}(y)$. 
We  refer to $K(x,y,t)$ as the \textit{heat kernel}. 
A standard proof is given in the Appendix.

\begin{figure}[htb]
 \begin{overpic} [width=\linewidth,trim={200 0 0 0},bb=0 0 1500 960]{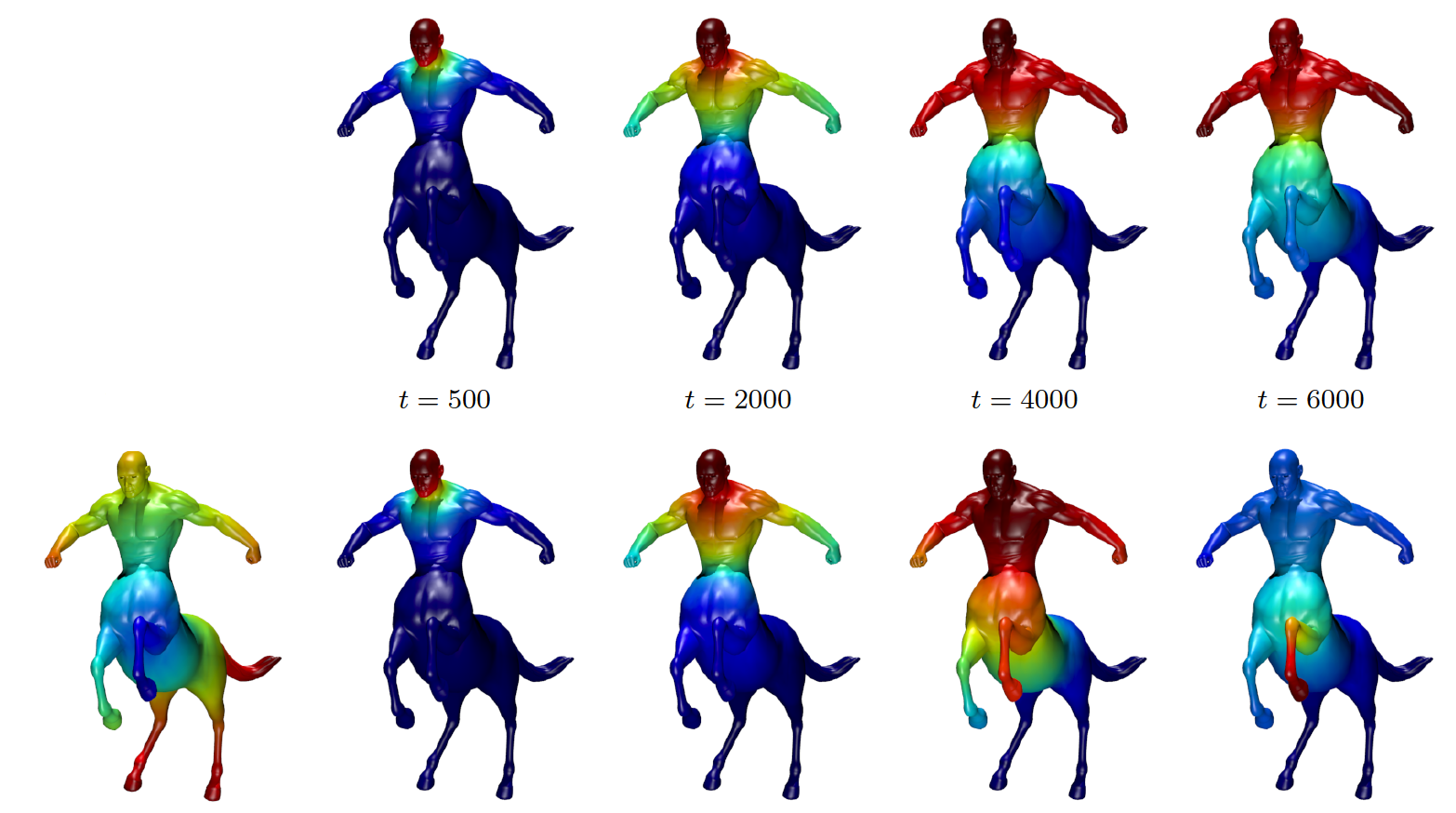}
 \put(23 ,67){\tiny{V}}
\end{overpic}
\centering
\caption{\label{fig:kernel}
Heat diffusion with a delta function at the centaur's head as initial condition. 
The diffusion is derived from the LBO (top) and the Hamiltonian (bottom) for different values of $t$. 
The potential $V$ used in this example is the geodesic distance from the front left leg. 
A signature extracted from a diffusion process using the Hamiltonian is more descriptive and in this case allows to resolve ambiguities due to symmetry.}
\end{figure}

According to the Feynman-Kac formula  \cite{feyman_kack}, the solution of the diffusion process is expressed in terms of the Wiener process,
\begin{equation}
\label{eq:feymanKack}
 u(x,t) = \mathbb{E}\Big(u_{0}(X_{t})\textrm{exp}\Big(\int_{0}^{t}V(X_{\tau})d\tau\Big)|X_{t}=x\Big).
\end{equation}
In the Laplacian case, the  initial value $u_{0}(x)$ is carried over random paths in time, while the expected value of the stochastic process is equal to the solution $u(x,t)$.
For $V>0$, the diffusion spreads according to the potential on the manifold, when the transported value is modulated exponentially by the potential $V$, diffusing anisotropically to low potential regions, as shown in Figure \ref{fig:kernel}.%
  
\subsection{Nodal sets}

An interesting property of the Laplacian is the relation between its eigenfunctions,  the number of connected nodal (zero) sets, and the number of complementary regions they define. 
Given an eigenfunction $\psi_i:\mathcal{M}\rightarrow \mathbb{R}$, a nodal set is defined as the set of points at which the eigenfunction values are zero. 
That is, 
\begin{equation}\label{eq:functionalRepresentation}
\mathcal{N}( \psi_{i} )=\{x\in \mathcal{M}|\psi_{i}(x)=0\}.
\end{equation}
The Nodal Theorem \cite{courant1966methods} states that the $i$-th eigenfunction of the LBO can split $\mathcal{M}$ to at most $i$ connected sub-domains.
In other words, the zero set of the $i$-th eigenfunction can separate the manifold into at most $i$ connected components. 
\begin{prop}
 \textit{
  Given the self-adjoint Hamiltonian operator $H$ on $\mathcal{M}$, with arbitrary boundary conditions; if its eigenfunctions are ordered  according to increasing eigenvalues, then,  the nodal set of the $i$-th eigenfunction divides the domain into no more than $i$ connected sub-domains.}
\end{prop}
The proof is essentially the same as that of the Laplacian case. 
 See \cite{courant1966methods} Vol.1 Sec. VI.6 for a proof.

\begin{figure}
\begin{overpic} [width=\linewidth,bb=0 0 1280 960]{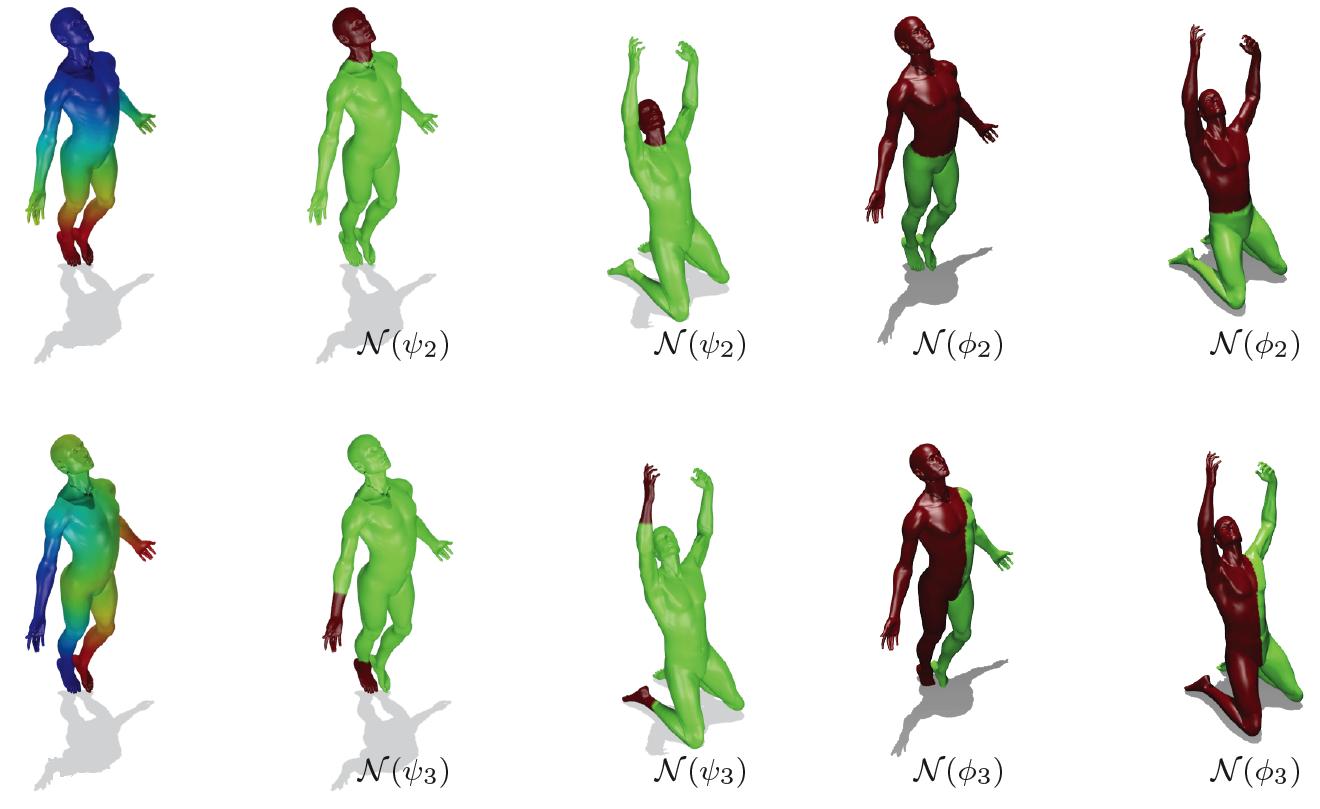}
\put(15,83){{V}}
\put(15,5){{V}}
\end{overpic}
 \centering
\caption{\label{fig:NodalSets} 
 Nodal domains obtained from the nodal sets of the Hamiltonian (second and third columns) and the LBO (fourth and fifth columns). Two potentials are depicted (first and second rows). 
 One can observe a meaningful segmentation induced by the nodal sets of the Hamiltonian. }
\end{figure}

As shown in Figure \ref{fig:finiteStep}, the Hamiltonian eigenfunctions  are tuned by the potential. 
Thus, shape segmentation can be obtained by separating the surface according to the induced nodal sets as described in \cite{levy2006laplace}.
Given a potential $V$ defined on the surface, meaningful segmentation can be induced by the nodal domains of the resulting eigenfunctions, as presented in  Figure \ref{fig:NodalSets}.

\subsection{Discretization}

In the discrete setting, we consider a triangular mesh $M$ in $\mathbb{R}^{3}$ with the associated space of functions that are continuous and linear in every triangle.
According to the Finite Element Method (FEM) \cite{dziuk1988finite}, the solution of the Hamiltonian eigenvalue problem (\ref{eq:sch_eigen}) can be computed by imposing that the equation $Hf = E f$ is satisfied in a weak sense, that is,
\begin{equation}\label{eq:fem1}
\begin{aligned}
\langle Hf,\varphi_j\rangle_{\mathcal{M}}=E \langle f,\varphi_j\rangle_{\mathcal{M}}, 
\end{aligned}
\end{equation} 
where $\varphi_{j}$ denote the Lagrange basis of piecewiselinear hat-functions on $M$, that take the value one at a vertex $m_{j}$ and vanishes at all the other vertices. $E$ represents the eigenenergy of the Hamiltonian. 
On the mesh, the bilinear form $\langle\cdot,\cdot\rangle_{{\mathcal{M}}}$ is evaluated by splitting the integrals into a sum over the triangles $T$ of $M$ by
\begin{equation}\label{eq:fem2}
\begin{aligned}
\langle u,v\rangle_{\mathcal{M}}=\sum_{T\in M}\langle u,v\rangle_{T}=\sum_{T\in M}\int_{T}uv da.
\end{aligned}
\end{equation}
Since the Hamiltonian is a linear operator we have
\begin{equation}\label{eq:fem3}
\begin{aligned}
\langle Hf,\varphi_j\rangle_{\mathcal{M}}=\langle -\Delta_{\mathcal{M}} f,\varphi_j\rangle_{\mathcal{M}}+\langle Vf,\varphi_j\rangle_{\mathcal{M}} . 
\end{aligned}
\end{equation} 
The matrix representation of $\langle -\Delta_{\mathcal{M}} f,\varphi_j\rangle_{\mathcal{M}}$ and $\lambda\langle f,\varphi_j\rangle_{\mathcal{M}}$ with respect to the Lagrange basis are well known \cite{pinkall1993computing} and define the stiffness matrix ${W}$ and the mass matrix ${A}$ with the entries
\begin{equation}\label{eq:fem4}
\begin{aligned}
 {W}_{ij}=\langle\nabla \varphi_{i},\nabla\varphi_{j}\rangle_{{\mathcal{M}}} \quad \text{and} 
 \quad {A}_{ij}=\langle\varphi_{i},\varphi_{j}\rangle_{{\mathcal{M}}}.
\end{aligned}
\end{equation} 
Thus, 
\begin{equation}\label{eq:fem5}
\begin{aligned}
\langle Vf,\varphi_j\rangle_{\mathcal{M}}&=\sum_{T\in M}\langle Vf,\varphi_j\rangle_{T}\\ &=\sum_{T\in M}\sum_{i}f_{i}\langle V\varphi_i,\varphi_j\rangle_{T}={AV}f,
\end{aligned}
\end{equation} 
where the last equality is obtained by representing the potential function $V$ as a diagonal matrix ${V}$ according to the Lagrange basis functions.
The discretization of the eigenvalue problem (\ref{eq:sch_eigen}) is defined by finding all pairs $\{E,\psi\}$ such that
\begin{equation}\label{eq:fem6}
\begin{aligned}
{H}\psi = {W}\psi+{A}{V}\psi=({W}+{A}{V})\psi=E {A}\psi.
\end{aligned}
\end{equation}
Efficient solution methods can be found in \cite{vallet2008spectral}.
Among the possible explicit representations of the matrices ${A}$ and ${W}$, we use here the cotangent formula \cite{pinkall1993computing,meyer2003discrete} where the stiffness matrix is defined as 
\begin{equation}\label{eq:cotan}
\begin{aligned}
 {W}_{ij} = \left 
  \{ \begin{array}{ll}
    -\sum_{j \neq i}{W}_{ij}, 						&  i = j, (i,j)\in N_i  \cr
  (\mbox{cot}\alpha_{ij}+\mbox{cot}\beta_{ij})/2,  & i \neq j, (i,j)\in N_i,
  \end{array} \right .
\end{aligned}
\end{equation}
 with $N_{i}=\{j:(i,j) \in \Gamma\}$, where $\Gamma$ is the set of edges of the triangulated surface interpreted as a graph and $\alpha_{ij},\beta_{ij}$ denote the angles
 $\angle ikj$ and $\angle jhi$ of the triangles sharing the edge $ij$. 
 The mass matrix is replaced by a diagonal lumped mass matrix of the area of local mixed Voronoi cells about each vertex $m_{i}$.
The manifold inner product is discretized as $\langle f,g\rangle_{A}=f^{T}Ag$.
Since ${V}$ only modifies the diagonal of ${W}$, our operator remains a sparse matrix with the same effective entries, and thus, there is no increase in the computational cost of the generalized eigendecomposition compared to that of the LBO.
 
\subsection{Robustness to noise}
\begin{figure}
\begin{overpic} [width=1\linewidth,trim=0 350 0 0,clip]{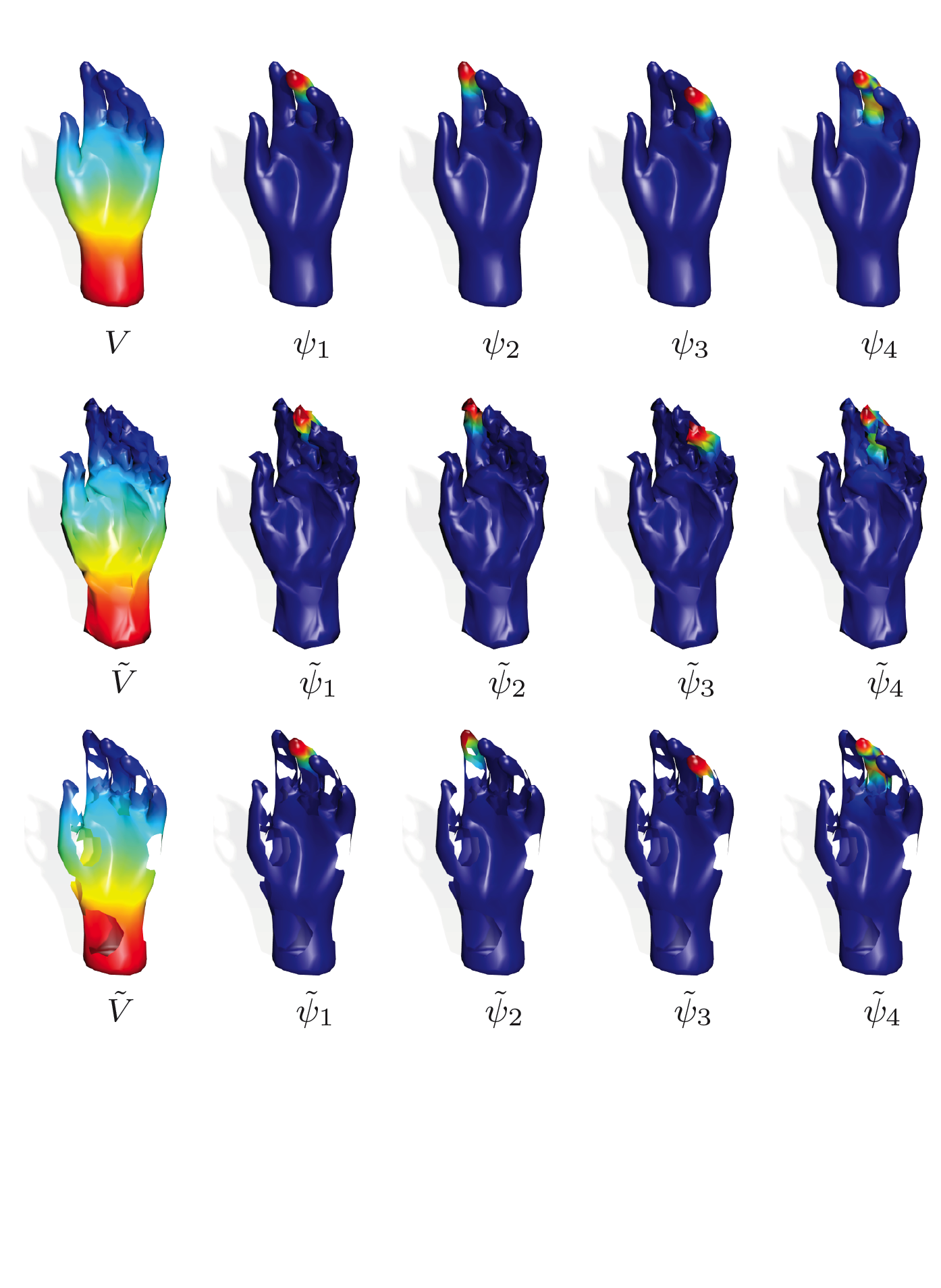}
\end{overpic}
 \centering
\caption{\label{fig:meshNoise} 
 Robustness to noise of the Hamiltonian. 
 First eigenfunctions $\psi_i$ of the Hamiltonian under potential $V$ (top). 
  First eigenfunctions $\tilde{\psi}_i$ of the Hamiltonian
  subject to Gaussian noise in positions of the vertices and 
  the potential (middle).
   First eigenfunctions $\tilde{\psi}_i$ of the Hamiltonian 
   subject topological noise (bottom). 
}
\end{figure}

As a generalization of the Laplacian, the Hamiltonian 
 exhibits similar robustness to noise. 
Consider the Hamiltonian matrix ${H}={A}^{-1}({W}+{AV})$ with $V$ the potential.
Then, the perturbed Hamiltonian has the form 
$\tilde{{H}}=\tilde{{A}}^{-1}(\tilde{{W}}+\tilde{{A}}\tilde{{V}})$. 
Let us define $\delta_{{A}} = |{A}-\tilde{{A}}|$ and $\delta_{{W}} = |({W}-\tilde{{W}})+({AV}-\tilde{{A}}\tilde{{V}})|$. 
Based on perturbation theory, and up to second-order corrections,
 the $i$-th eigenfunction $\tilde{\psi}_{i}$ of $\tilde{H}$ has the form  
\begin{equation}\label{eq:perturbation}
\begin{aligned}
\tilde{\psi}_{i} =\psi_{i}(1-\frac{\psi_{i}^{T}\delta_{A}\psi_{i}}{2})+\sum_{k\neq i}\frac{\psi_{i}^{T}(\delta_{W}-{E}_{i}\delta_{A})\psi_{k}}{{E}_{i}-{E}_{k}}\psi_{k},
\end{aligned}
\end{equation} 
with $\psi_{i}$ and $E_{i}$ being respectively the $i$-th eigenfunction and eigenvalue of the unperturbed Hamiltonian.
Assuming uniformly distributed random noise on the mesh, the eigenfunctions of the regular Laplacian 
 may present smaller distortion to noise than the Hamiltonian since the perturbation 
  is amplified by area and potential distortions. 
Still, in case of potential with small values the distortion is insignificant.
In Figure \ref{fig:meshNoise}, we present the original surface and its noisy version in which vertex positions have been corrupted by additive Gaussian noise with $\sigma_{x}^{2}=20\%$ of the mean edge length. 
The potential is also modified by adding  a Gaussian noise with $\sigma_{V}^{2}=20\%$ 
 of the initial variance of the potential.

The construction of the Laplacian depends crucially on the mesh connectivity making
 it sensitive to topological noise such as holes and part removal that can be found 
in many depth acquisition scenarios.
The compact support of the basis elements of the Hamiltonian 
 makes it  robust to noise compared to the basis elements 
 that are generated by the Laplacian. 
We illustrate the robustness property in Figure
 \ref{fig:meshNoise} where $30\%$ of the 
 surface area was removed due to topological noise in the
 form of small holes.

\section{Optimization of the potential}
\label{section:optimization_label}
One natural problem emerging when working with the Hamiltonian is the ability to define an optimal potential function for a specific task.
The choice of the potential is application dependent but can be represented through minimization problem generically defined as
\begin{equation}\label{eq:PCA_H_0}
\begin{aligned}
& \underset{V}{\text{min}}
& & D(X,V) \\
& \text{s.t.}
& & V\in \mathbb{R}^{n},\\
\end{aligned}
\end{equation}
where $D(X,V)$ denotes the data term depending on the data matrix $X$ and the vector $V$ defining the diagonal potential matrix. 
Regularization terms can be further be added. If the analytical solution remains complex, a common approach is to minimize the goal function with an optimization algorithm involving the gradient of the goal function with respect to the potential. In this section we propose an optimization framework based on perturbation theory of the eigenvectors where optimal potential is obtained.
To that end, we need to derive the gradient $\nabla_{V}D$ for a given objective $D$.

Here we will consider the problem of data representation using the discrete basis of the Hamiltonian referred to as $\Psi_{k}(V)=\Psi_{k}\in \mathbb{R}^{n\times k}$ representing the $k$ eigenvectors
of the Hamiltonian such that $\Psi_{k}^{T}\Psi_{k}=I_{k}$. The discretized minimization problem is defined as 
\begin{equation}\label{eq:PCA_H}
\begin{aligned}
& \underset{V}{\text{min}}
& & \|\Psi_{k}\Psi_{k}^{T}X-X\|_{F}^{2} \\
& \text{s.t.}
& & V\in \mathbb{R}^{n},\\
\end{aligned}
\end{equation}
with $k<n$. The objective defines the representation error of the data $X$ in the subspace spanned by the columns of  $\Psi_{k}$ and in the sense of the Frobenius norm $\|\cdot\|_{F}$. 
For a general orthonormal matrix $\Psi_{k}$, the problem is equivalent to Principal Component Analysis (PCA). We can straightforwardly obtain that 
\begin{equation}\label{eq:PCA_H_1}
\begin{aligned}
L&= \|\Psi_{k}\Psi_{k}^{T}X-X\|_{F}^{2} =\text{trace}\big((\Psi_{k}\Psi_{k}^{T}X-X)^{T}(\Psi_{k}\Psi_{k}^{T}X-X)\big) \\
& =\text{trace}\big(X^{T}X\big)+ \text{trace}\big(X^{T}\Psi_{k}\Psi^{T}\Psi\Psi_{k}^{T}X\big) \\ & \ \  -2\text{trace}\big(X^{T}\Psi_{k}\Psi_{k}^{T}X\big) 
\\
& =-\text{trace}\big(\Psi_{k}\Psi_{k}^{T}XX^{T}\big)+\text{trace}\big(XX^{T}\big).
\end{aligned}
\end{equation}
Thus, the differential of the loss function $L$ with respect to $V$ is obtained by
\begin{equation}\label{eq:PCA_H_2}
\begin{aligned}
dL&= -d\text{trace} \big(\Psi_{k}\Psi_{k}^{T}XX^{T}\big)\\
&= -\text{trace} \big(d\Psi_{k}\Psi_{k}^{T}XX^{T}\big)-\text{trace} \big(\Psi_{k} d\Psi_{k}^{T}XX^{T}\big)\\
&= -2\text{trace} \big(\Psi_{k}^{T}XX^{T}d\Psi_{k}\big).
\end{aligned}
\end{equation}
It remains to derive the differential of the Hamiltonian eigenvectors.
Let us consider the full matrix of eigenvectors $\Psi_{n}\in\mathbb{R}^{n\times n}$, the $n \times n$ diagonal matrix of eigenenergies $[\Lambda]_{ii}=\lambda_{i}$ and the discrete Hamiltonian operator $H$. The eigenvalue decomposition problem is given by $H\Psi_{n}=\Psi_{n}\Lambda$.
Thus, the differential of the spectral decomposition problem is given by 
\begin{equation}\label{eq:PCA_H_3}
\begin{aligned}
dH\Psi_{n}+Hd\Psi_{k}=d\Psi_{n}\Lambda+\Psi_{n}d\Lambda.
\end{aligned}
\end{equation}
Multiplying by $\Psi_{n}^{T}$ on the left side and denoting $d\Psi_{n}=\Psi_{n}C$ with $C\in \mathbb{R}^{n\times n}$, we have
\begin{equation}\label{eq:PCA_H_4}
\begin{aligned}
&\Psi_{n}^{T}dH \Psi_{n}+\Psi_{n}^{T}H\Psi_{n}C=\Psi_{n}^{T}\Psi_{n}C\Lambda+\Psi_{n}^{T}\Psi_{n}d\Lambda \\
 & \Psi_{n}^{T}dH \Psi_{n}+\Lambda C=C\Lambda+d\Lambda,
\end{aligned}
\end{equation}
since $\Psi_{n}^{T}\Psi_{n}=I_{n}$.
We readily obtain that the off diagonal elements of the matrix $C$ can be defined by
\begin{equation}\label{eq:PCA_H_5}
\begin{aligned}
C_{ij} = \frac{(\Psi^{i})^{T}dH\Psi^{j}}{\lambda_{j}-\lambda_{i}} &, \forall i\neq j.
\end{aligned}
\end{equation}
Here $\Psi^{j}$ represents the $j$-th column of the matrix of eigenvectors.
The diagonal elements of $C$ are defined by the following 
\begin{equation}\label{eq:PCA_H_6}
\begin{aligned}
(\Psi_{n}+d\Psi_{n})^{T}(\Psi_{n}+d\Psi_{n}) &=& I\\
 \Psi_{n}^{T}\Psi_{n}+\Psi_{n}^{T}d\Psi_{n}+d\Psi_{n}^{T}\Psi_{n}+d\Psi_{n}^{T}d\Psi_{n} &=& I\\
 I+\Psi_{n}^{T}\Psi_{n}C+C^{T}\Psi_{n}^{T}\Psi_{n}+C^{T}\Psi_{n}^{T}\Psi_{n}C &=& I\\
  C+C^{T}+C^{T}C &=& 0.
\end{aligned}
\end{equation}
The diagonal elements are then defined by $2C_{ii}+\sum_{k=1}^{n}C_{ki}^{2}=0$.
Since second order elements are negligible, we have $C_{ii} = 0$. 
We obtain that $d\Psi_{n}=\Psi_{n}C=\Psi_{n}(\Psi_{n}^{T}dH\Psi_{n})\odot B$, with $\odot$ denoting the Hadamard product and the matrix $B$ defined as
\begin{equation}\label{eq:PCA_H_7}
\begin{aligned}
B_{ij}=\left \{
  \begin{tabular}{cc}
  $\frac{1}{\lambda_j-\lambda_i}$ &, $i\neq j$  \\~\\
  0 &, $i=j$.
  \end{tabular}
\right.
\end{aligned}
\end{equation}

The selection of the first $k$ eigenvectors $d\Psi_{k}$ are obtained by multiplying $d\Psi_{n}$ by the truncated identity matrix $Z=I_{n\times k}$. The differential is now known and can be plugged into (\ref{eq:PCA_H_2}) in order to extract $dH$, that is:
\begin{equation}
\begin{aligned}
dL&= -2\text{trace} \big(\Psi_{k}^{T}XX^{T}d\Psi_{k}\big)\\
&= -2\text{trace} \big(\Psi_{k}^{T}XX^{T}\Psi_{n}CZ\big)\\
&= -2\text{trace} \big(\Psi_{k}^{T}XX^{T}\Psi_{n}\big(\Psi_{n}^{T}dH\Psi_{n}\odot B\big)Z\big)\\
&= -2\text{trace} \big(\Psi_{n}\big(Z\Psi_{k}^{T}XX^{T}\Psi_{n}\big)\odot B\Psi_{n}^{T}dH\big)\\
&= \langle\big(-2\big(\Psi_{n}\big(Z\Psi_{k}^{T}XX^{T}\Psi_{n}\big)\odot B\Psi_{n}^{T}\big)^{T},dH\rangle.
\end{aligned}
\end{equation}
The passage in the fourth line stems from the equivalence $\text{trace}(A(B\odot C))=\text{trace}((A\odot C^{T})B)$.
Since $dH = d(L+\text{diag}(V))= \text{diag}(dV)$, we obtain finally 
\begin{equation}
\begin{aligned}
\nabla_{V}L=\text{diag}\big(-2\big(\Psi_{n}\big(Z\Psi_{k}^{T}XX^{T}\Psi_{n}\big)\odot B\Psi_{n}^{T}\big)\big).
\end{aligned}
\end{equation}

Two problems arise from the suggested scheme. First, the high computational cost of a full (sparse) matrix diagonalization.
Second, the matrix $C$ remains undefined when eigenvectors have non-trivial multiplicities.
The first problem can be relaxed by approximating the matrix $d\Psi$ with less eigenvectors. This is especially justified for distant indices, where the eigenenergies are well separated and the corresponding elements of matrix $B$ become negligible. 
Also, the data can be projected onto the LBO basis so the solution complexity remains constant with the size of the mesh.
Even if the second problem has been treated in \cite{van2007computation}, it seems that lack of smoothness at isolated points is not critical for computation and convergence by resorting to a sub-gradient approach. The alternative opted for here is to stabilize the matrix $B$ in order to avoid exploding gradients. We use the approximation
\begin{equation}
\begin{aligned}
B_{ij}\approx\frac{1}{(|\lambda_j-\lambda_i|+\epsilon)(\text{sign}(\lambda_j-\lambda_i))},
\end{aligned}
\end{equation}
where the $\text{sign}$ function is not vanishing.

In its geometric setting, the solution remains similar and is obtained by defining a new basis $\tilde{\Psi}_{k}=A^{\frac{1}{2}}\Psi_{k}$, such that $\tilde{\Psi}_{k}^{T}\tilde{\Psi}_{k}=I_{k}$, coupled with the consistently discretized Frobenius norm $\|X\|_{A}^{2}=\langle X,X \rangle_{A}=\|A^{\frac{1}{2}}X\|_{F}^{2}$.
In the following experiments we allowed negative potential for performance consideration only, since the potential is defined over the whole codomain $\mathbb{R}$.
Also, for physically interpretable solutions we enforced positive potential by using quadratic function $V^2$. The extension of the derivation is straightforward but decreased the performance since it is more restrictive.  

\begin{figure}[htbp]
\begin{overpic}[width=1\linewidth,trim={90 30 90 0},clip]{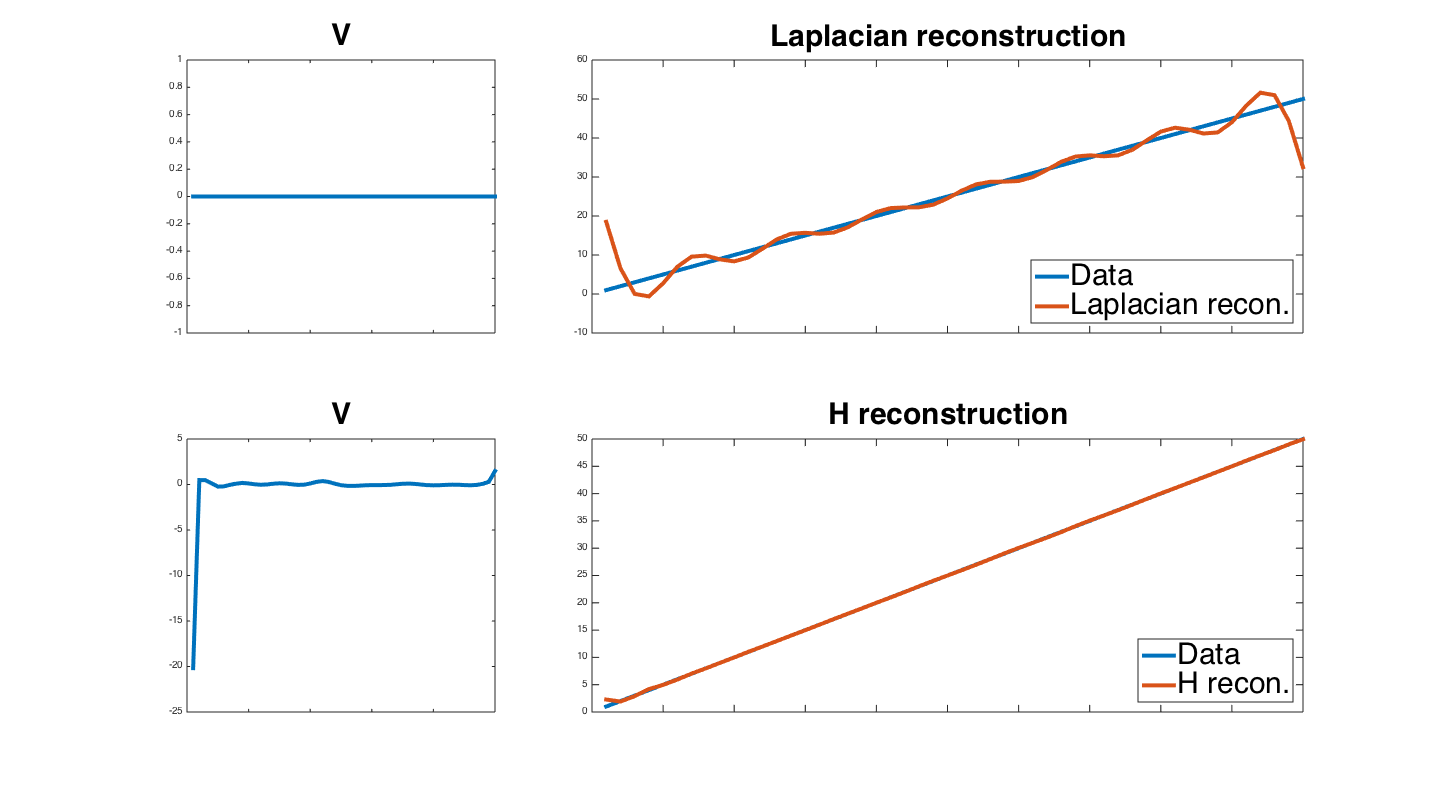}
\centering
 \end{overpic}
 \caption{\label{fig:euclideanPCA0} 
Reconstruction of a linear function using the Laplacian and the Hamiltonian constructed with the proposed framework. 15 eigenvectors were used in this experiment. Observe that the potential is high close to the boundary to reduce the representation error.}
 \end{figure}

\begin{figure}[htbp]
\begin{overpic}[width=1\linewidth,bb=0 0 1280 960,trim={0 0 0 0},clip]{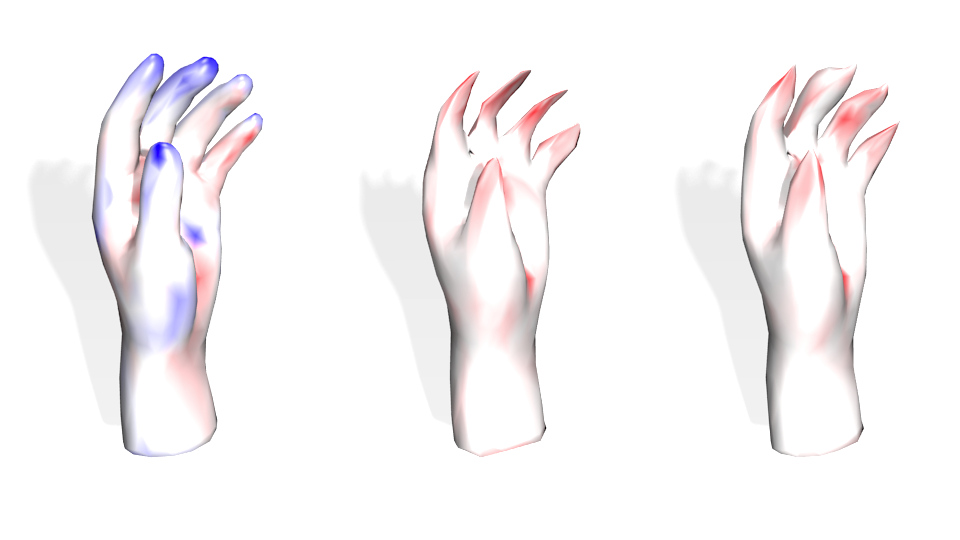}
 \put (38,10) {$V$}
 \put (117,10) {LBO}
 \put (185,10) {Hamiltonian}

\centering
 \end{overpic}
 \caption{\label{fig:manifolPCA} 
Potential function defined on the original mesh (left), reconstruction of the mesh coordinates with 50 eigenvectors using the LBO (middle) and the Hamiltonian constructed with the proposed method (right). Blue and red colors represent negative and positive values respectively. The Hamiltonian is able to focus on sharp regions of the mesh designated by the blue regions of the potential for a better reconstruction (fingers). The errors are 0.0015 and 0.00061 for the LBO and the Hamiltonian respectively. 
}
 \end{figure}

As a toy experiment, we propose to find the best potential for the representation of a function in the one dimensional Euclidean domain. Given a function $f\in \mathbb{R}^{n}$, we seek for the best potential minimizing $\|\Psi_{k}\Psi_{k}^{T}f-f\|_{2}^{2}$. 
We compare in Figure \ref{fig:euclideanPCA0} the reconstruction performance on a one dimensional linear function with the Laplacian and the Hamiltonian built from the optimized potential.
In Figure \ref{fig:manifolPCA}, we propose to reconstruct the matrix of coordinates of a mesh so the data matrix is defined by $X=(x,y,z) \in \mathbb{R}^{n\times 3}$.
The experiments were implemented using the quasi-Newton method with initial zero potential. 

\begin{figure}[htbp]
\begin{overpic}[width=1\linewidth,trim={40 0 20 0},clip]{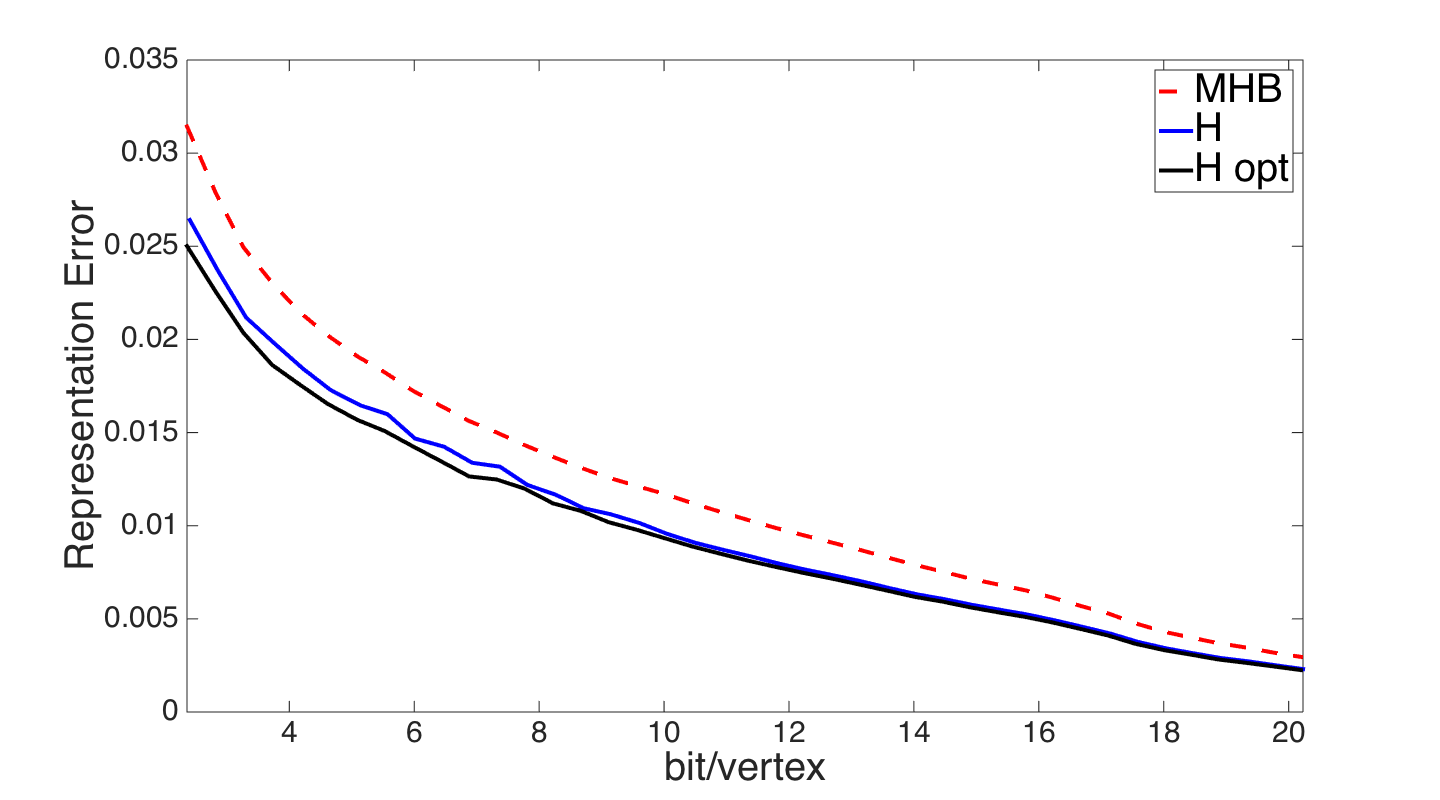} 
\centering
\end{overpic}
\caption{\label{fig:compression_Res} 
 Geometry compression performance comparison between the Laplacian (MHB), the proposed projected operator (H) and the optimal Hamiltonian (H opt) using the proposed framework for the sharp Fandisk shape. The optimal Hamiltonian performance are presented with no encoding of the potential itself.  }
 \end{figure}

An important application related data representation is spectral mesh compression. \cite{karni2000spectral} proposed to project the coordinates functions of the mesh onto the LBO eigenfunctions in order to encode the mesh geometry via the first coefficients only. Since most of the function energy is generally contained in the first coefficients, the reconstruction distortion is low, up to fine details related to higher frequencies. Since matrix decomposition is an expensive operation, they suggested to segment the shape into smaller parts that can be processed separately. By sending the mesh topology (triangles) separately, the combinatorial graph Laplacian is built on the decoder side and the signal can be reconstructed with the received coefficients.
We suggest to apply this idea to our basis which potential $V$ is obtained by the proposed optimization framework. 
However, one major drawback is that we need to encode the potential as well as the coefficients.
Also, some methods use the ordering of the vertices in order to encode information \cite{touma}.
Here we suggest to reorder the vertices such that the vertex with the smallest potential is be assigned the index $1$ and the vertex with the largest potential is be assigned the index $n$.
By using a fixed potential defined as
\begin{equation}
\tilde{V}=\text{diag}(1,...,n),
\end{equation}
the decoder simply applies $L+\alpha \tilde{V} +\beta$ in order to obtain the Hamiltonian basis.
Here $\alpha$ and $\beta$ are the regression coefficients minimizing $\|\alpha \tilde{V}+\beta-V\|_{2}^{2}$ that are also encoded. We present compression results in Figure \ref{fig:compression_Res}. 

\section{Compressed Manifold Modes}
\label{section:compressed_modes}
\cite{Ozolins2013} proposed a a novel method to create a set of localized eigenfunctions in Euclidean domains. 
To that end, they modified the construction of standard differential operators by adding an $L_1$ regularization term to the variational leading to the decomposition of the operator. The resulting eigenfunctions were called compressed modes and were shown to be compactly supported \cite{brezis1974solutions}.
\cite{neumann2014compressed} extended this construction to manifolds, suggesting the following discrete $L_1$ regularization problem
\begin{equation}\label{eq:CMM-discrete}
\begin{aligned}
& \underset{{\Phi}}{\text{min}}
& & \operatorname{trace} ( {\Phi}^\mathrm{T} {W} {\Phi} ) + \mu \| {\Phi} \|_1 \ \\
& \text{s.t.}
& & {\Phi}^\mathrm{T} {A} {\Phi} = {I}.
\end{aligned}
\end{equation}
with the parameter $\mu$ that controlling the localization of the basis. 
Proposed solutions require the use of expensive and unstable optimization techniques (\cite{neumann2014compressed,kovnatsky2015madmm}) based on ADMM and proximal operators.

The latter optimization problem (\ref{eq:CMM-discrete}) can be written as an Hamiltonian eigendecomposition problem \cite{Bron_Chouk_Kim_Sel}
\begin{equation}\label{eq:CMM-wl2}
\begin{aligned}
& \underset{{\Phi}}{\text{min}}
& & \operatorname{trace} ( {\Phi}^\mathrm{T} {W} {\Phi} ) + \mu \operatorname{trace}(\Phi^{T}V_{i}\Phi) \ \\
& \text{s.t.}
& & {\Phi}^\mathrm{T} {A} {\Phi} = {I},
\end{aligned}
\end{equation}
where $V_i$ is the diagonal matrix operator defining the potential that corresponds to the $i$-th eigenvector that localizes the support of $\phi_i$ in low-potential areas. The potential is defined iteratively using a reweighted least squares scheme
\begin{equation}\label{eq:CMM-ls}
\begin{aligned}
V_{i}&=\frac{1}{2|\phi_{i}|},
\end{aligned}
\end{equation}
ensuring that the minimizers of (\ref{eq:CMM-discrete}) and (\ref{eq:CMM-wl2}) coincide.
Interestingly, the potential here is defined as a function of the eigenfunction, namely $V_{i}=V_{i}(\phi_{i})$. The potential and the resulting eigenstate are then intrinsically linked, meaning that the potential is influenced by the state of the particle itself. Consequently, a perturbation of the potential enforces perturbation of the eigenfunction and vice versa until reaching steady state.

We formulate the compressed manifold modes problem as
\begin{equation}\label{eq:CMM-discrete_}
\begin{aligned}
& \underset{{\phi}_i}{\text{min}}
& & {\phi}_i^\mathrm{T} H_{i}  {\phi}_i + \beta \sum_{j < i} \| {\phi}_j^\mathrm{T} {A} {\phi}_i  \|_2^2 \\
& \text{s.t.}
& & {\phi}_i^\mathrm{T}{A} {\phi}_i = 1,
\end{aligned}
\end{equation}
with ${H}_i = W+\mu A V_{i}$ and where $\beta$ is a sufficiently large constant such that the third term guarantees that the $i$-th mode ${\phi}_i$ is ${A}$-orthogonal to the previously computed modes ${\phi}_j$, $j<i$. Observe that albeit non-convex, the problem has a closed form global solution, that is the smallest generalized eigenvector ${\phi}_i$ of
\begin{equation}
(H_{i} + {Z}_i) {\phi}_i = \lambda_i {A} {\phi}_i
\label{eq:eigendec-irls}
\end{equation}
with
$$
{Z}_i =U_{i}U_{i}^{T}=\beta {A}\left( \sum_{j<i} {\phi}_j {\phi}_j^\mathrm{T} \right){A}.
$$
For small number of compressed modes, $\bb{Z}_i$ is a low rank matrix and finding the smallest generalized eigenvector can be solved efficiently since the involved matrix is the sum of a sparse and a low-rank matrix.

\begin{figure}[t]
\centering
        \includegraphics[width=0.9\linewidth,trim={0 0 0 0},clip]{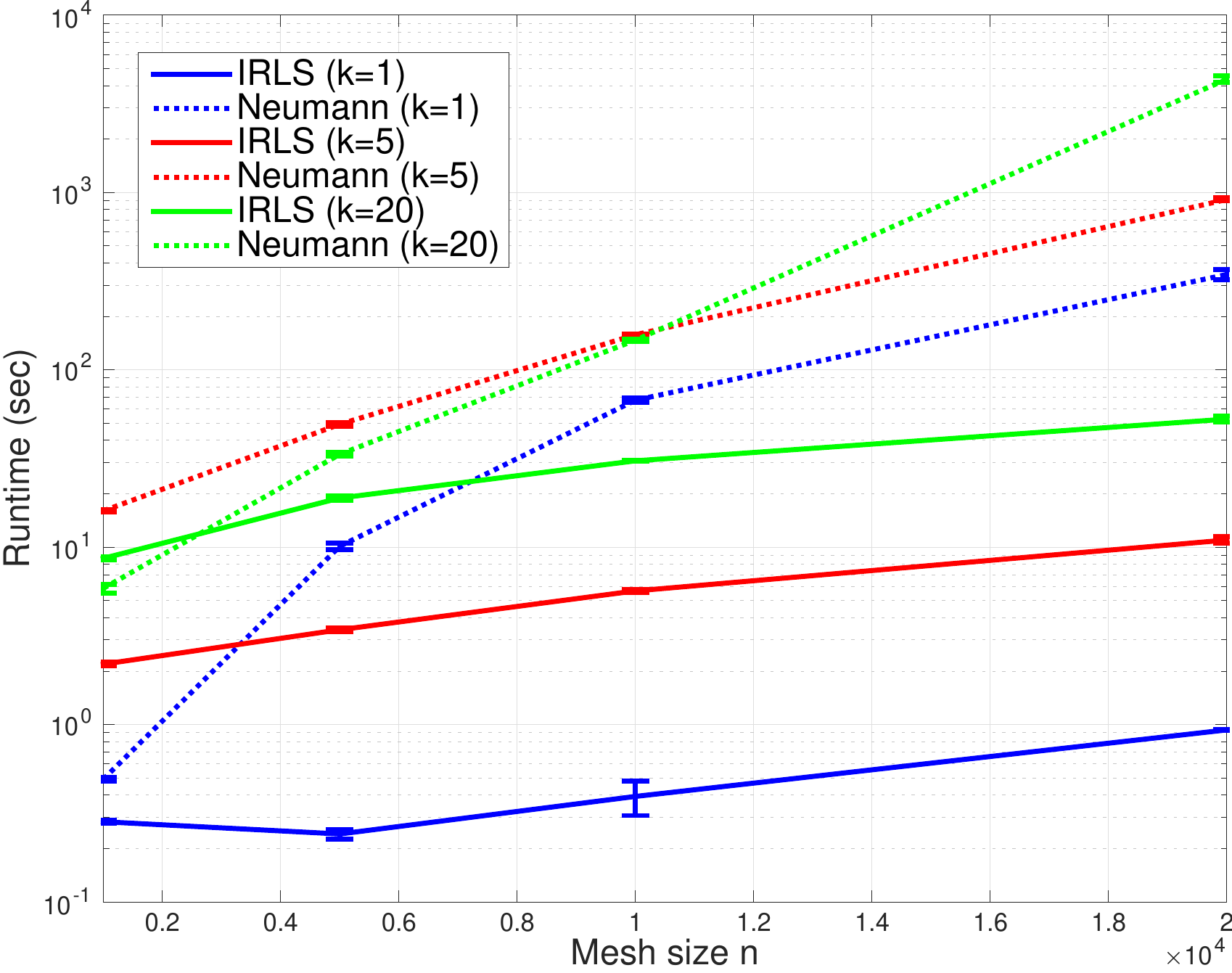}\\
 \caption{\label{fig:runtimes} Runtimes of Neumann et al. and the proposed framework on meshes of varying size (number of vertices $n$) and number of eigenvectors $k$. Averages and standard deviations are presented over $10$ runs. Same stopping criteria were applied to all methods.}
\end{figure}

Several numerical eigendecomposition implementations use the Arnoldi iteration algorithm.
In our matrix decomposition problem, the core operation is the multiplication by the inverse of the matrix with a vector, operation that cannot be solved straightforwardly. Also, shifting the maximum eigenvalue using power method is too unstable since it depend on gap of the first eigenvalues, generally tight.
In our configuration the Woodbury identity
$$
({H}_{i}+{U}_{i}{U}_{i}^\mathrm{T})^{-1} = {H}_{i}^{-1} - {H}_{i}^{-1} {U}_{i} ( {I} + {U}_{i}^\mathrm{T} {H}_{i}^{-1}{U}_{i} )^{-1} {U}_{i}^\mathrm{T} {H}_{i}^{-1}
$$
can be used to compute efficiently the vector multiplication with the inverse of the matrix as a cascade of sparse and low-rank systems.
Unlike solutions of the inconsistently discretized problem \ref{eq:CMM-discrete}, the basis obtained with the proposed Hamiltonian framework is more robust under various discretizations and can be computed at a fraction of the computational cost (Figure \ref{fig:runtimes}).  

\section{Shape matching}
\label{section:Shape_matching}
The task of matching pairs of shapes lies at the core of many shape analysis tasks and plays a central role in operations such as 3D alignment and shape reconstruction. 
While rigid shape matching has been well studied in the literature, non-rigid correspondence remains a difficult task even for nearly isometric surfaces. 
When dealing with rigid objects, it is sufficient to find the rotation and translation that aligns one shape to the other \cite{chen1991object}. 
Therefore, the rigid matching problem amounts to determining only six degrees of freedom. 
At the other end, non-rigid matching generally requires dealing with many more degrees of freedom.
Since the LBO is invariant to isometric deformations, it has been used extensively to aid the solution of correspondence problem.
Several properties of the Hamiltonian operator make it a better choice for this task compared to its zero-potential particular case that is the LBO.  \\
 \noindent{\textbf{Invariance.}} 
The Laplace-Beltrami Operator is defined in terms of 
 the metric tensor which is invariant to isometries. 
For a potential function defined intrinsically, the resulting Hamiltonian is also isometry-invariant.
 
\noindent{\textbf{Compactness.}}
Compactness means that scalar functions on a shape should be well approximated by using only a small
 number of basis elements.
From Theorem \ref{H_Optimality} and as a generalization of the Laplacian, the global support and 
  compactness hold for a bounded (low) potential.

\noindent{\textbf{Descriptiveness.}}
The LBO eigenvalues are related to frequency.  
Similarly, eigenenergies of the Hamiltonian relate to the number of oscillations on the manifold.
Theorem \ref{proofEigenvalues} demonstrates that the modes corresponding to small eigenvalues of the Hamiltonian defined with 
 a positive potential, encapsulate higher frequencies, even when localized, compared to the modes of the regular LBO. 
At the other end, highly oscillating eigenfunctions can be used to represent fine details of the shape that can be crucial for shape matching. 
Also, the potential enforces different oscillations in different regions on the manifold, allowing for better discrimination of similar areas and disambiguation of intrinsic symmetries with asymmetric potential.

\begin{figure}[htbp]
\centering
\subfigure[Nearly isometric shapes]{
     \begin{tikzpicture}
    \node (HKS_TOSCA) {\includegraphics[scale=0.15,bb=0 0 1280 960,trim={250 50 60 50},clip]{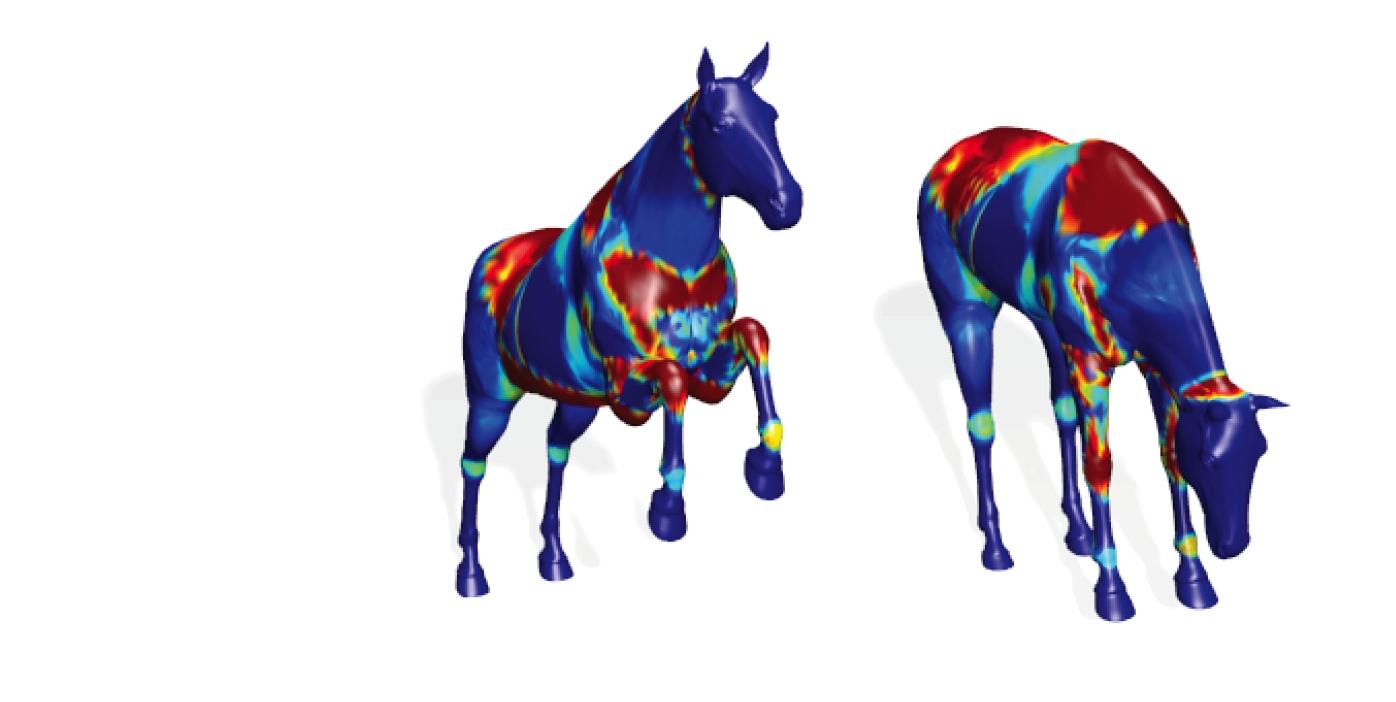}};
    \end{tikzpicture}}
  \quad
 \subfigure[LBO]{
     \begin{tikzpicture}
    \node (HKS_SCAPE) {\includegraphics[scale=0.2,trim={70 30 70 0},clip]{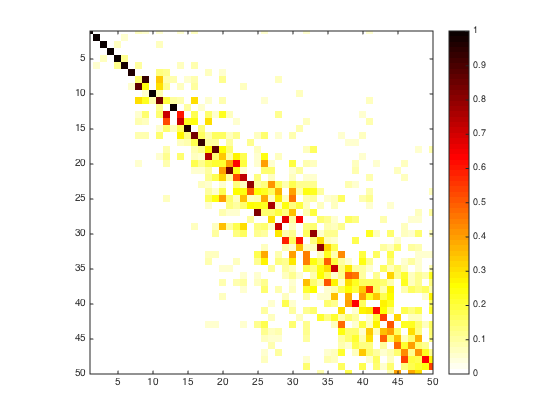}     \put(-55,75){\scriptsize{{  }}}};
    \end{tikzpicture}
    }
      \quad
 \subfigure[Hamiltonian]{
     \begin{tikzpicture}
    \node (HKS_SCAPE) {\includegraphics[scale=0.2,trim={70 30 70 0},clip]{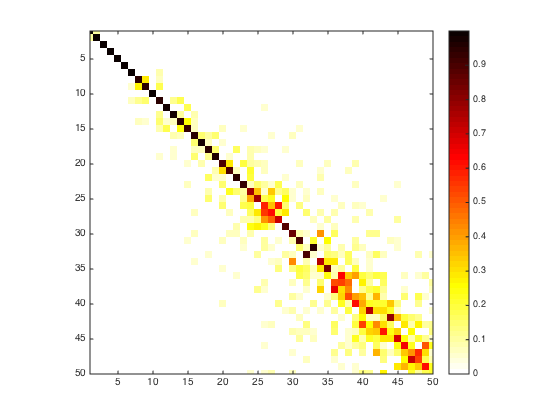}     \put(-55,75){\scriptsize{{   }}}};
    \end{tikzpicture}
    }
\caption{
 \label{fig:C_Matrix}
Two nearly isometric meshes with high potential (hot colors) in large
 distortion regions (a), 
 functional maps matrix $C$ of the LBO (b) and the Hamiltonian (c). 
 }
\end{figure}

\noindent{\textbf{Stability}.}
Deformations of non-rigid shapes and articulated objects 
 can stretch the surface. 
In such cases, the LBO eigendecomposition of the two   
 shapes will be different. 
We could compensate for such local metric distortions  
 by carefully designing a potential.
Assigning high potential to strongly distorted regions 
 would lead to lower values of the eigenfunctions in those areas (\ref{eq:SchFunctional}).
Such a potential will reduce the discrepancy between 
 corresponding eigenfunctions at least for the
 lower eigenergies, as shown via the functional maps representations \cite{ovsjanikov2012functional} in Figure \ref{fig:C_Matrix}.
Let define $A_{{M}}(m_i)$ and $A_{{N}}(n_i)$, the area at vertex $m_i$ on mesh ${M}$ and $n_i$ on the second mesh ${N}$ respectively and $\tau:{M}\rightarrow {N}$ a bijection between two (discretized) surfaces ${M}$ and ${N}$.
Then, we define the potential $V$ at vertex $m_i=\tau^{-1}(n_{i})$ as 
 \begin{equation}\label{eq:stablePotential}
V(m_i) = \text{max}\Bigg\{\frac{A_{{M}}(m_i)}{A_{{N}}(n_i)}, \frac{A_{{N}}(n_i)}{A_{{M}}(m_i)}\Bigg\}.
\end{equation}
%

Among the few stable intrinsic invariants that can be extracted from the geometry,  we will use the stable first eigenfunctions of the LBO and geodesic distances. 
Additional non necessarily intrinsic  information such as photometric properties or even extrinsic shape properties such as principal curvatures \cite{hildebrandt2012modal} can also be integrated into the potential field. 

\subsection{Experimental Evaluation}
\label{sec:results}
We tested the proposed basis and compared its matching performances
 to that the LBO basis as applied to pairs of triangulated meshes of shapes
 from the TOSCA dataset \cite{bronstein2008numerical} and the SCAPE dataset
 \cite{dragomir2005correlated}.
The TOSCA data set contains densely sampled synthetic human and animal surfaces, 
 divided into several classes with given ground-truth point-to-point correspondences
  between the shapes within each class. 
The SCAPE data set contains scans of real human bodies in different poses. 
The evaluation method used is described in \cite{kim2011blended} where the distortion
 curves describe the percentage of surface points falling within a relative geodesic
 distance from what is assumed to be their true locations.
Symmetries were not allowed in all evaluations. 
Note that we assume that the sign ambiguity of the first eigenfunctions generating the potential
 is resolved \cite{shtern2013matching}. 

\begin{figure}[htb]
\centering
\subfigure[TOSCA]{
     \begin{tikzpicture}
    \node (HKS_TOSCA) 
     {\includegraphics[scale=0.21,trim={0 0 0 0},clip]
                      {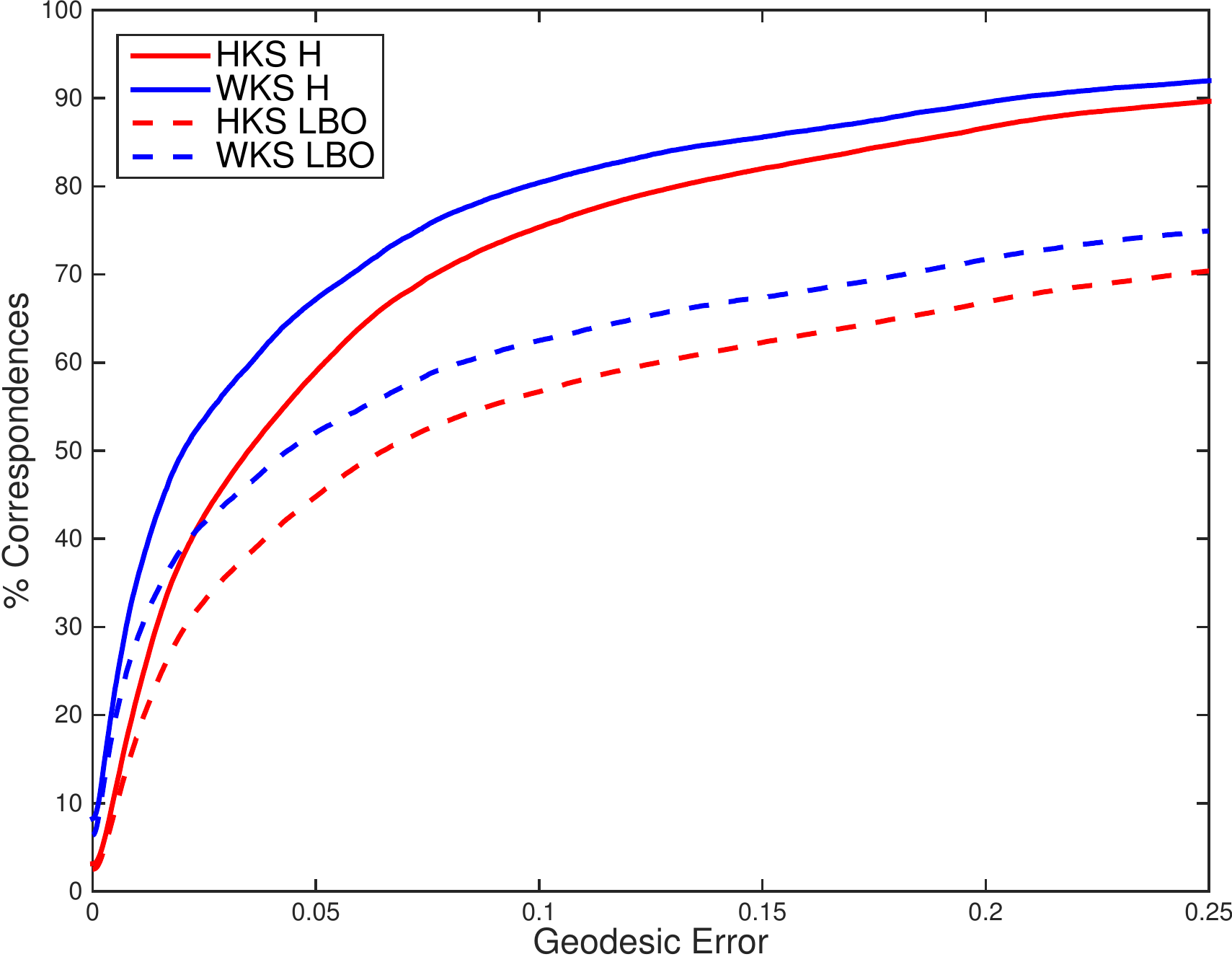}
    \put(-75,80){\scriptsize{\textbf{ }}}};
    \end{tikzpicture}
}
  \quad
 \subfigure[SCAPE]{

     \begin{tikzpicture}
    \node (HKS_SCAPE) {\includegraphics[scale=0.21,trim={0 0 0 0},clip]{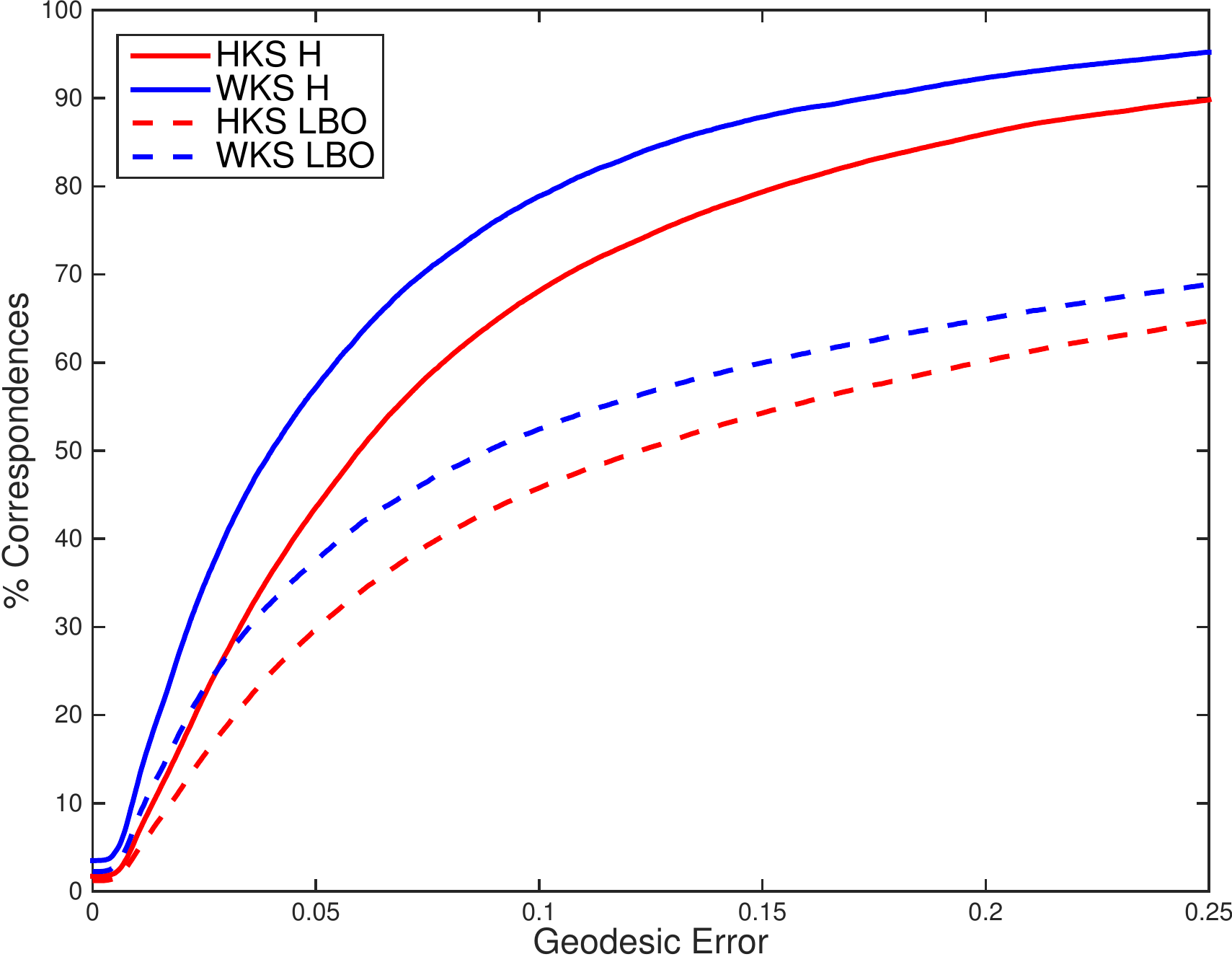}
    \put(-75,80){\scriptsize{\textbf{ }}}};
    \end{tikzpicture}
    }
\caption{
 \label{fig:HKSreg}
  Evaluation of the diffusion kernels signatures matches on the TOSCA and SCAPE datasets.}
\end{figure}

Figure \ref{fig:HKSreg} compares the two operators by matching  diffusion kernel descriptors 
derived from the corresponding eigenfunctions. 
The diffusion on the shape using the Hamiltonian as the diffusion operator is more descriptive than regular diffusion that cannot resolve the symmetries. 
Also, it would be natural to compute the WKS
signature when the \SCH equation is governed by a given effective potential. 
As intrinsic positive potential we use the normalized sum of the four first nontrivial eigenfunctions of the LBO on each shape, adding a constant of minimal value in order to obtain a non-negative potential. 
This way only the intrinsic unstable geometry of the shape is involved in defining the Hamiltonian operator. 

\begin{figure}[htb]
\centering
\subfigure[TOSCA]{
     \begin{tikzpicture}
    \node (Area_TOSCA) {\includegraphics[scale=0.21,trim={0 0 0 0},clip]{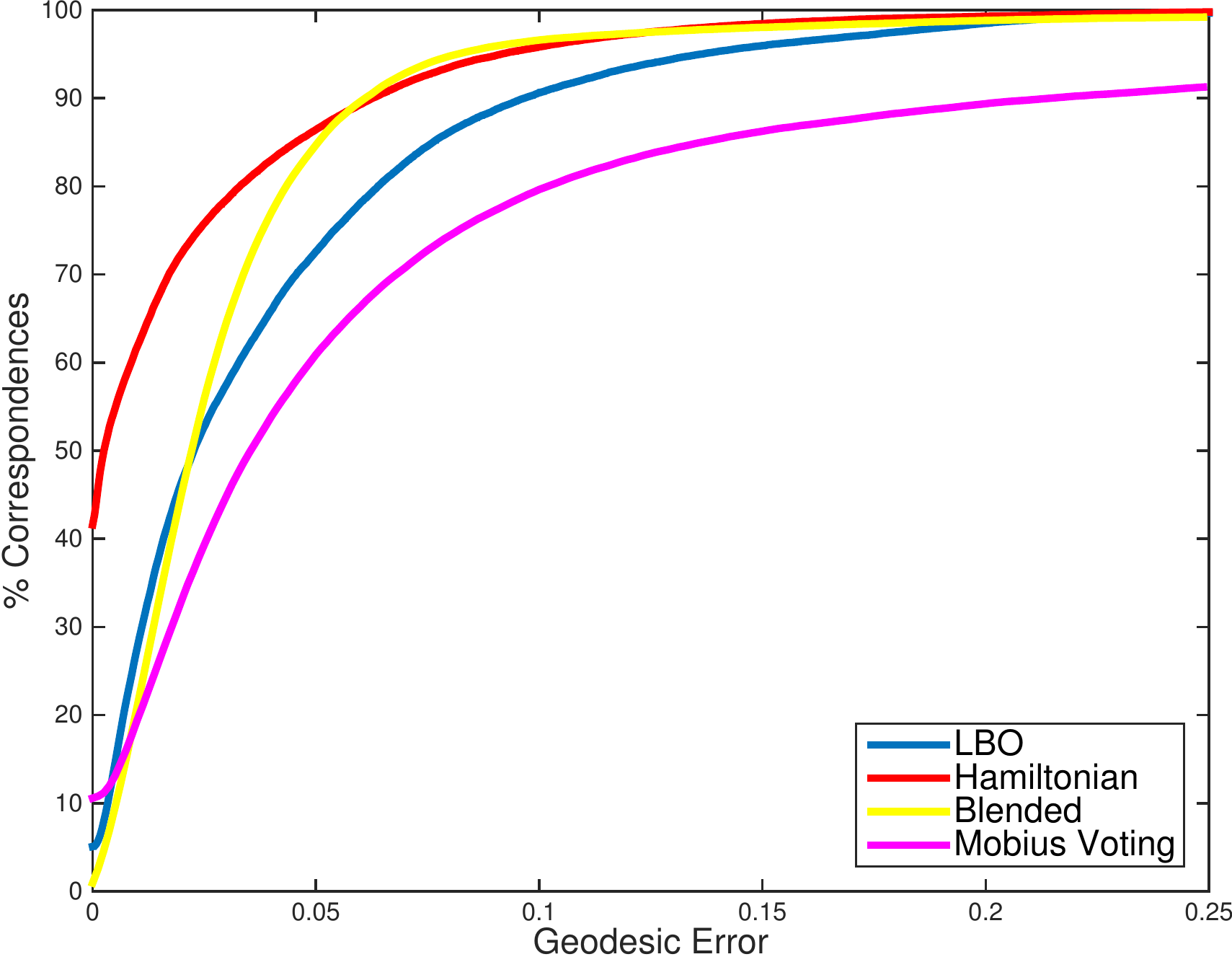}
};
    \end{tikzpicture}
}
  \quad
 \subfigure[SCAPE]{

     \begin{tikzpicture}
    \node (Area_SCAPE) {\includegraphics[scale=0.21,trim={0 0 0 0},clip]{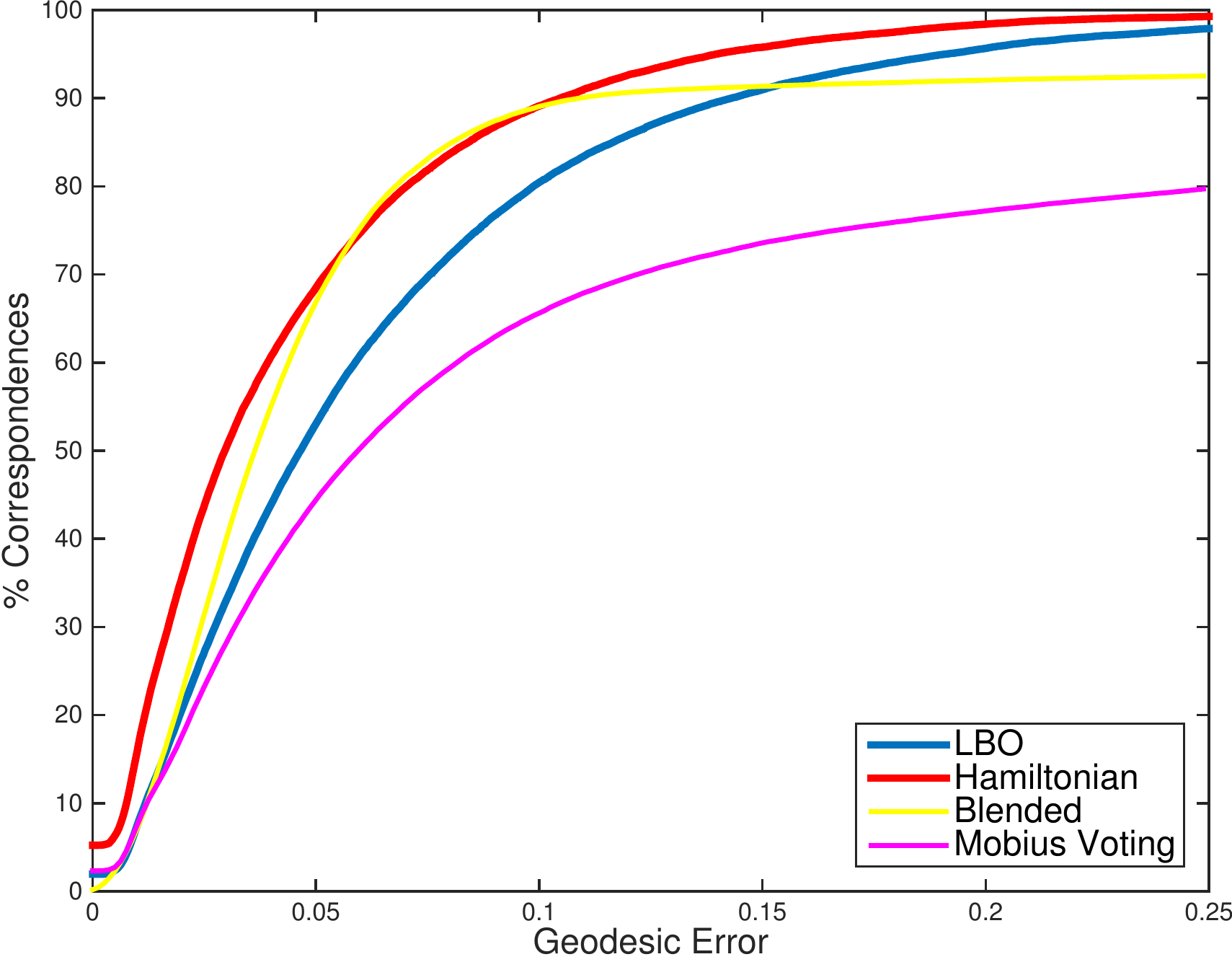}
};
    \end{tikzpicture}
    }
\caption{
 \label{fig:bench_area}
  Evaluation of the spectral signature matches on 
   the TOSCA and SCAPE data-sets.}
\end{figure}

In case we know which regions are prone to elastic distortions, like joints and stretchable skin in articulated objects, we could  suppress the effect of those regions in our matching procedures by using an appropriate potential as a selective mask. 
Figure \ref{fig:bench_area}, compares the operator with and without potential by matching the spectral signatures computed by the framework of \cite{shtern2015spectral}. 
The potential we used is the local area distortion when comparing the meshes of two corresponding objects, as in (\ref{eq:stablePotential}). 
The descriptiveness of the potential and the localization of the harmonics  lead to more accurate matching results. 

\begin{figure}[htb]
\centering
\subfigure[Photometric data]{
 \begin{overpic}[scale=0.2,trim={250 60 250 0},clip]{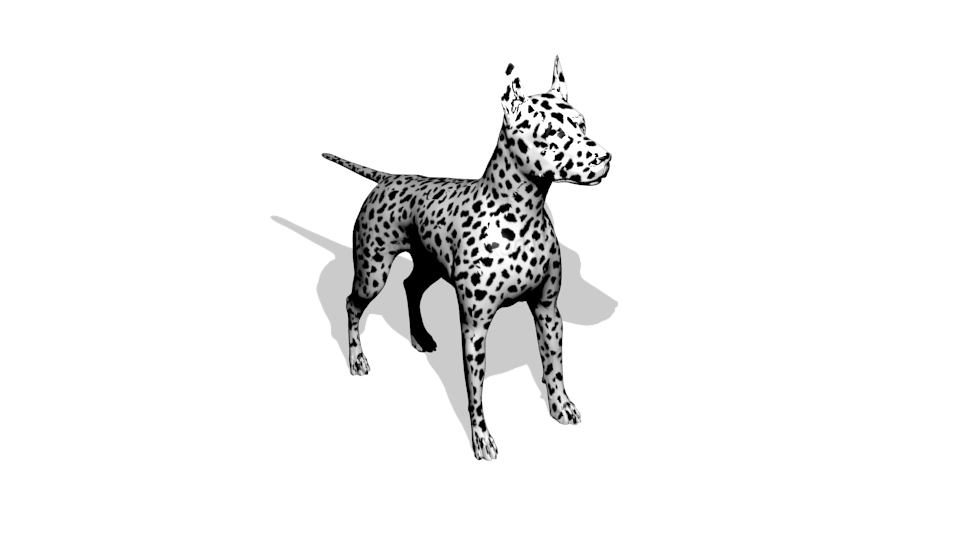}
 \end{overpic}
}
 \quad
 \subfigure[Signatures]{
 \begin{overpic}[scale=0.21,trim={0 0 0 0},clip]{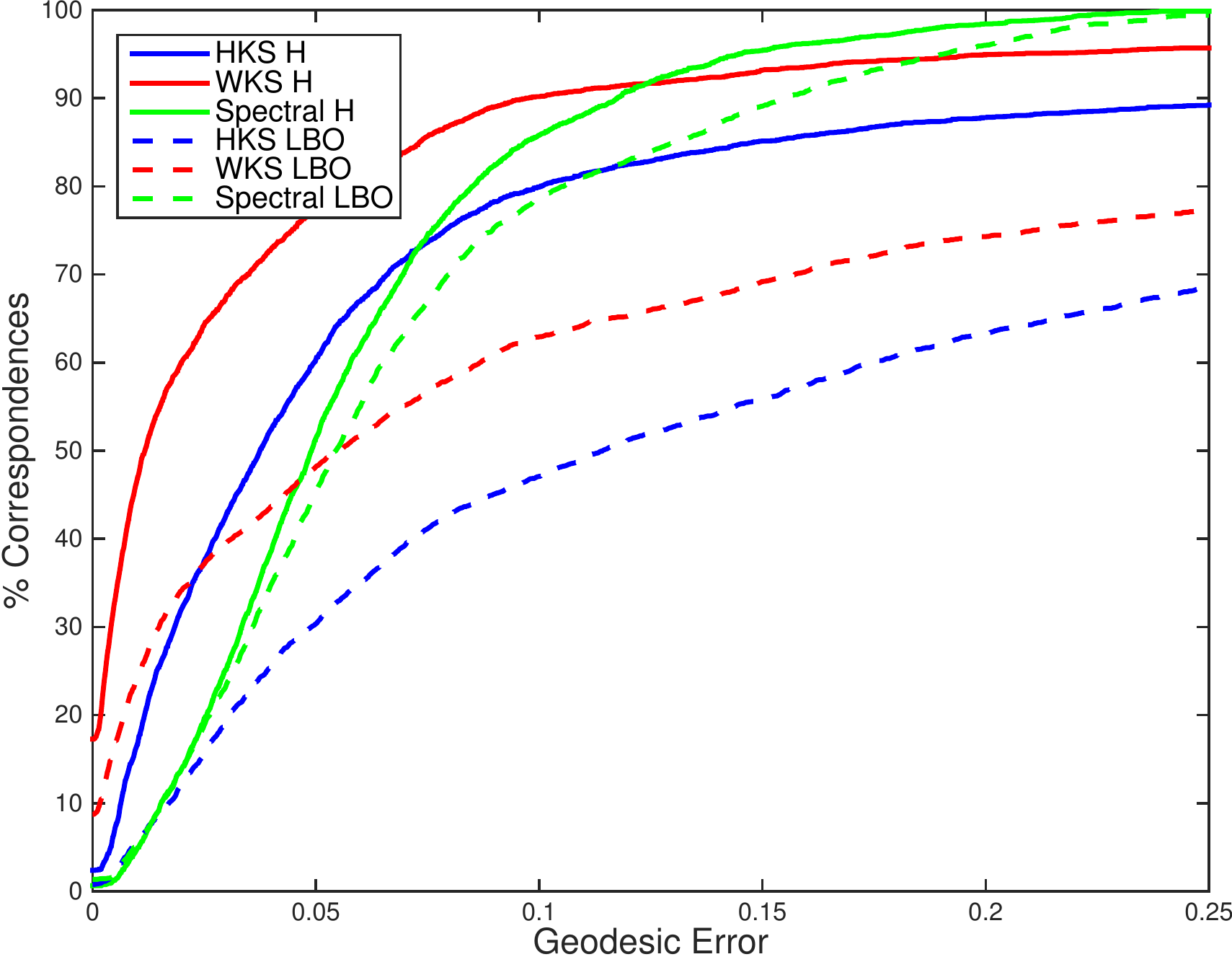}
 \end{overpic} 
 } 
\caption{\label{fig:dalmatian}
 Evaluation of the descriptors matches on the "Dogs" benchmark from the TOSCA dataset with dalmatian texture.}
\end{figure}

To investigate the performances of the Hamiltonian with
 photometric textures used as potential,  we present in 
 Figure \ref{fig:dalmatian} the results of different 
 signatures matching with a dalmatian texture defined 
 for the "Dogs" shapes from the TOSCA data set.
 
\begin{figure}[htb]
\centering
\begin{tabular}{l r}
 \begin{overpic}[scale=0.3]{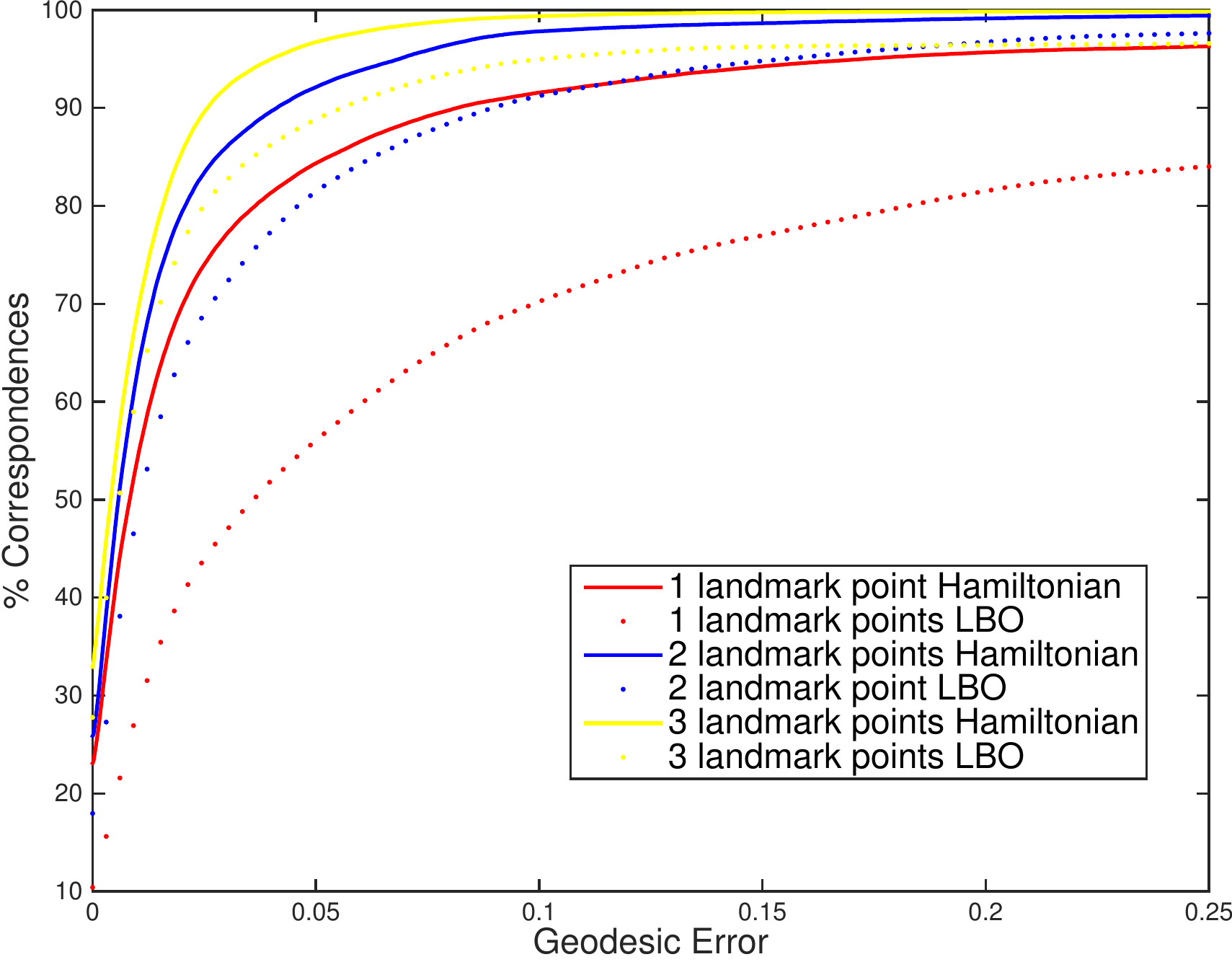}
 \put(13,120){\scriptsize{\textbf{ }}}
\end{overpic}&
\end{tabular}
\caption{\label{fig:initPointsGraphs}Evaluation of the 
 ICSKM algorithm with different landmark initialization 
 matches on the TOSCA dataset. 
 We used geodesic distances from  given landmark points 
 as intrinsic geometric potential on the shapes.}
\end{figure}

Iterative refinement of functional representations have been proven to be powerful in shape matching \cite{ovsjanikov2012functional}.
Given an initial partial or dense map, it tries to recover iteratively dense and accurate matching between two given shapes. 
Here we use a similar refinement framework dubbed as Iterative Closest Spectral Kernel Maps (ICSKM) \cite{shterniterative} for performance comparison between the two bases.    
Figure \ref{fig:initPointsGraphs} compares the regular ICSKM algorithm working with the Laplacian eigenspace and the Hamiltonian method when we provided one, two, or three landmark points,  that were randomly selected from the  ground-truth mapping. 
The potential used in these examples is the geodesic distance from the landmark points.
Note that again we use only the geometry of the shapes in order to refine the match between them using the new basis. 

%
\section{Conclusion}
A classical operator was adopted from the field of quantum mechanics and  adapted to shape analysis problems. 
Functional and spectral properties of the Hamiltonian operator were presented and compared to the popular Laplacian operator often used in many shape analysis procedures.
A general optimization method for solving variational problems involving the Hamiltonian operator have been proposed and employed to the task of mesh compression.
Features and texture properties can be incorporated into the new operator to obtain a descriptive and stable basis that provides a powerful domain of operation for shape matching. 
Various directions for future research include exploration of the operator on other shape analysis tasks such as partial shape matching where occluded areas could be refined via the
potential. 
\begin{acks}
This work has been supported by Grant agreement No. 267414 of the European Community’s FP7-ERC program.
\end{acks}

\bibliographystyle{acmtog}
\bibliography{references.bib}

\begin{thebibliography}{}

\bibitem[\protect\citeauthoryear{Aflalo, Brezis, and Kimmel}{Aflalo
  et~al\mbox{.}}{2015}]{aflalo2014optimality}
{\sc Aflalo, Y.}, {\sc Brezis, H.}, {\sc and} {\sc Kimmel, R.} 2015.
\newblock On the optimality of shape and data representation in the spectral
  domain.
\newblock {\em {SIAM} J. Imaging Sciences\/}~{\em 8,\/}~2, 1141--1160.

\bibitem[\protect\citeauthoryear{Aflalo, Dubrovina, and Kimmel}{Aflalo
  et~al\mbox{.}}{2016}]{aflalo2013spectral}
{\sc Aflalo, Y.}, {\sc Dubrovina, A.}, {\sc and} {\sc Kimmel, R.} 2016.
\newblock Spectral generalized multi-dimensional scaling.
\newblock {\em International Journal of Computer Vision (IJCV)\/}~{\em
  118,\/}~3, 380--392.

\bibitem[\protect\citeauthoryear{Anguelov, Srinivasan, Pang, Koller, Thrun, and
  Davis}{Anguelov et~al\mbox{.}}{2004}]{dragomir2005correlated}
{\sc Anguelov, D.}, {\sc Srinivasan, P.}, {\sc Pang, H.-C.}, {\sc Koller, D.},
  {\sc Thrun, S.}, {\sc and} {\sc Davis, J.} 2004.
\newblock The correlated correspondence algorithm for unsupervised registration
  of nonrigid surfaces.
\newblock In {\em Proceedings of the 17th International Conference on Neural
  Information Processing Systems}. NIPS'04. MIT Press, Cambridge, MA, USA,
  33--40.

\bibitem[\protect\citeauthoryear{Aubry, Schlickewei, and Cremers}{Aubry
  et~al\mbox{.}}{2011}]{aubry2011wave}
{\sc Aubry, M.}, {\sc Schlickewei, U.}, {\sc and} {\sc Cremers, D.} 2011.
\newblock The wave kernel signature: A quantum mechanical approach to shape
  analysis.
\newblock In {\em Computer Vision Workshops (ICCV Workshops), 2011 IEEE
  International Conference on}. 1626--1633.

\bibitem[\protect\citeauthoryear{Boscaini, Masci, Rodol\`a, and
  Bronstein}{Boscaini et~al\mbox{.}}{2016}]{BosMasRodBro16}
{\sc Boscaini, D.}, {\sc Masci, J.}, {\sc Rodol\`a, E.}, {\sc and} {\sc
  Bronstein, M.~M.} 2016.
\newblock Learning shape correspondence with anisotropic convolutional neural
  networks.
\newblock Tech. Rep. arXiv:1605.06437.

\bibitem[\protect\citeauthoryear{Brezis}{Brezis}{1974}]{brezis1974solutions}
{\sc Brezis, H.} 1974.
\newblock Solutions with compact support of variational inequalities.
\newblock {\em Russian Mathematical Surveys\/}~{\em 29,\/}~2, 103--108.

\bibitem[\protect\citeauthoryear{Brezis}{Brezis}{2010}]{brezis2010functional}
{\sc Brezis, H.} 2010.
\newblock {\em Functional analysis, Sobolev spaces and partial differential
  equations}.
\newblock Springer Science \& Business Media.

\bibitem[\protect\citeauthoryear{Bronstein, Bronstein, and Kimmel}{Bronstein
  et~al\mbox{.}}{2008}]{bronstein2008numerical}
{\sc Bronstein, A.}, {\sc Bronstein, M.}, {\sc and} {\sc Kimmel, R.} 2008.
\newblock {\em Numerical geometry of non-rigid shapes\/}, 1 ed.
\newblock Springer Publishing Company, Incorporated.

\bibitem[\protect\citeauthoryear{{Bronstein}, {Choukroun}, {Kimmel}, and
  {Sela}}{{Bronstein} et~al\mbox{.}}{2016}]{Bron_Chouk_Kim_Sel}
{\sc {Bronstein}, A.}, {\sc {Choukroun}, Y.}, {\sc {Kimmel}, R.}, {\sc and}
  {\sc {Sela}, M.} 2016.
\newblock "{Consistent discretization and minimization of the L1 norm on
  manifolds}".
\newblock In {\em 3D Vision (3DV), 2016 4nd International Conference on}. IEEE.

\bibitem[\protect\citeauthoryear{Bronstein, Bronstein, Guibas, and
  Ovsjanikov}{Bronstein et~al\mbox{.}}{2011}]{bronstein2011shape}
{\sc Bronstein, A.~M.}, {\sc Bronstein, M.~M.}, {\sc Guibas, L.~J.}, {\sc and}
  {\sc Ovsjanikov, M.} 2011.
\newblock Shape google: Geometric words and expressions for invariant shape
  retrieval.
\newblock {\em ACM Trans. Graph.\/}~{\em 30,\/}~1 (Feb.), 1:1--1:20.

\bibitem[\protect\citeauthoryear{Bronstein, Bronstein, and Kimmel}{Bronstein
  et~al\mbox{.}}{2006a}]{bronstein2006efficient}
{\sc Bronstein, A.~M.}, {\sc Bronstein, M.~M.}, {\sc and} {\sc Kimmel, R.}
  2006a.
\newblock Efficient computation of isometry-invariant distances between
  surfaces.
\newblock {\em {SIAM} J. Scientific Computing\/}~{\em 28,\/}~5, 1812--1836.

\bibitem[\protect\citeauthoryear{Bronstein, Bronstein, and Kimmel}{Bronstein
  et~al\mbox{.}}{2006b}]{bronstein2006generalized}
{\sc Bronstein, A.~M.}, {\sc Bronstein, M.~M.}, {\sc and} {\sc Kimmel, R.}
  2006b.
\newblock Generalized multidimensional scaling: A framework for
  isometry-invariant partial surface matching.
\newblock {\em Proceedings of the National Academy of Sciences\/}~{\em
  103,\/}~5, 1168--1172.

\bibitem[\protect\citeauthoryear{Chen and Koltun}{Chen and
  Koltun}{2015}]{chen2015robust}
{\sc Chen, Q.} {\sc and} {\sc Koltun, V.} 2015.
\newblock Robust nonrigid registration by convex optimization.
\newblock In {\em 2015 IEEE International Conference on Computer Vision
  (ICCV)}. 2039--2047.

\bibitem[\protect\citeauthoryear{Chen and Medioni}{Chen and
  Medioni}{1992}]{chen1991object}
{\sc Chen, Y.} {\sc and} {\sc Medioni, G.} 1992.
\newblock Object modelling by registration of multiple range images.
\newblock {\em Image Vision Comput.\/}~{\em 10,\/}~3 (Apr.), 145--155.

\bibitem[\protect\citeauthoryear{Courant and Hilbert}{Courant and
  Hilbert}{1966}]{courant1966methods}
{\sc Courant, R.} {\sc and} {\sc Hilbert, D.} 1966.
\newblock {\em Methods of mathematical physics}. Vol.~1.
\newblock CUP Archive.

\bibitem[\protect\citeauthoryear{Cox and Cox}{Cox and
  Cox}{2008}]{cox2000multidimensional}
{\sc Cox, A. M.~A.} {\sc and} {\sc Cox, F.~T.} 2008.
\newblock {\em Multidimensional Scaling}.
\newblock Springer Berlin Heidelberg, Berlin, Heidelberg, 315--347.

\bibitem[\protect\citeauthoryear{Dziuk}{Dziuk}{1988}]{dziuk1988finite}
{\sc Dziuk, G.} 1988.
\newblock Finite elements for the beltrami operator on arbitrary surfaces.
\newblock In {\em Partial differential equations and calculus of variations}.
  Springer, 142--155.

\bibitem[\protect\citeauthoryear{Elad and Kimmel}{Elad and
  Kimmel}{2003}]{elad2003bending}
{\sc Elad, A.} {\sc and} {\sc Kimmel, R.} 2003.
\newblock On bending invariant signatures for surfaces.
\newblock {\em IEEE Trans. Pattern Anal. Mach. Intell.\/}~{\em 25,\/}~10
  (Oct.), 1285--1295.

\bibitem[\protect\citeauthoryear{Griffiths}{Griffiths}{2005}]{griffiths2005introduction}
{\sc Griffiths, D.} 2005.
\newblock {\em Introduction to Quantum Mechanics}.
\newblock Pearson international edition. Pearson Prentice Hall.

\bibitem[\protect\citeauthoryear{Grossmann, Kiryati, and Kimmel}{Grossmann
  et~al\mbox{.}}{2002}]{grossmann2002computational}
{\sc Grossmann, R.}, {\sc Kiryati, N.}, {\sc and} {\sc Kimmel, R.} 2002.
\newblock Computational surface flattening: a voxel-based approach.
\newblock {\em IEEE Transactions on Pattern Analysis and Machine
  Intelligence\/}~{\em 24,\/}~4 (Apr), 433--441.

\bibitem[\protect\citeauthoryear{Hildebrandt, Schulz, von Tycowicz, and
  Polthier}{Hildebrandt et~al\mbox{.}}{2012}]{hildebrandt2012modal}
{\sc Hildebrandt, K.}, {\sc Schulz, C.}, {\sc von Tycowicz, C.}, {\sc and} {\sc
  Polthier, K.} 2012.
\newblock Modal shape analysis beyond laplacian.
\newblock {\em Computer Aided Geometric Design\/}~{\em 29,\/}~5, 204--218.

\bibitem[\protect\citeauthoryear{Iglesias and Kimmel}{Iglesias and
  Kimmel}{2012}]{iglesias2012schrodinger}
{\sc Iglesias, J.~A.} {\sc and} {\sc Kimmel, R.} 2012.
\newblock {\em Schr{\"o}dinger Diffusion for Shape Analysis with Texture}.
\newblock Springer Berlin Heidelberg, Berlin, Heidelberg, 123--132.

\bibitem[\protect\citeauthoryear{Karni and Gotsman}{Karni and
  Gotsman}{2000}]{karni2000spectral}
{\sc Karni, Z.} {\sc and} {\sc Gotsman, C.} 2000.
\newblock Spectral compression of mesh geometry.
\newblock In {\em Proceedings of the 27th Annual Conference on Computer
  Graphics and Interactive Techniques}. SIGGRAPH '00. ACM Press/Addison-Wesley
  Publishing Co., New York, NY, USA, 279--286.

\bibitem[\protect\citeauthoryear{Kim, Lipman, and Funkhouser}{Kim
  et~al\mbox{.}}{2011}]{kim2011blended}
{\sc Kim, V.~G.}, {\sc Lipman, Y.}, {\sc and} {\sc Funkhouser, T.} 2011.
\newblock Blended intrinsic maps.
\newblock {\em ACM Trans. Graph.\/}~{\em 30,\/}~4 (July), 79:1--79:12.

\bibitem[\protect\citeauthoryear{Kovnatsky, Bronstein, Bronstein, and
  Kimmel}{Kovnatsky et~al\mbox{.}}{2011}]{kovnatsky2011diffusion}
{\sc Kovnatsky, A.}, {\sc Bronstein, M.~M.}, {\sc Bronstein, A.~M.}, {\sc and}
  {\sc Kimmel, R.} 2011.
\newblock Photometric heat kernel signatures.
\newblock In {\em Scale Space and Variational Methods in Computer Vision -
  Third International Conference, {SSVM} 2011, Ein-Gedi, Israel, May 29 - June
  2, 2011, Revised Selected Papers}. 616--627.

\bibitem[\protect\citeauthoryear{Kovnatsky, Glashoff, and Bronstein}{Kovnatsky
  et~al\mbox{.}}{2016}]{kovnatsky2015madmm}
{\sc Kovnatsky, A.}, {\sc Glashoff, K.}, {\sc and} {\sc Bronstein, M.~M.} 2016.
\newblock {MADMM}: a generic algorithm for non-smooth optimization on
  manifolds.
\newblock In {\em European Conference on Computer Vision}. Springer, 680--696.

\bibitem[\protect\citeauthoryear{Levy}{Levy}{2006}]{levy2006laplace}
{\sc Levy, B.} 2006.
\newblock Laplace-beltrami eigenfunctions towards an algorithm that
  "understands" geometry.
\newblock In {\em Proceedings of the IEEE International Conference on Shape
  Modeling and Applications 2006}. SMI '06. IEEE Computer Society, Washington,
  DC, USA, 13--.

\bibitem[\protect\citeauthoryear{Litman and Bronstein}{Litman and
  Bronstein}{2014}]{litman2014learning}
{\sc Litman, R.} {\sc and} {\sc Bronstein, A.~M.} 2014.
\newblock Learning spectral descriptors for deformable shape correspondence.
\newblock {\em IEEE transactions on pattern analysis and machine
  intelligence\/}~{\em 36,\/}~1, 171--180.

\bibitem[\protect\citeauthoryear{M{\'e}moli and Sapiro}{M{\'e}moli and
  Sapiro}{2005}]{memoli2005theoretical}
{\sc M{\'e}moli, F.} {\sc and} {\sc Sapiro, G.} 2005.
\newblock A theoretical and computational framework for isometry invariant
  recognition of point cloud data.
\newblock {\em Foundations of Computational Mathematics\/}~{\em 5,\/}~3,
  313--347.

\bibitem[\protect\citeauthoryear{Meyer, Desbrun, Schr{\"o}der, and Barr}{Meyer
  et~al\mbox{.}}{2003}]{meyer2003discrete}
{\sc Meyer, M.}, {\sc Desbrun, M.}, {\sc Schr{\"o}der, P.}, {\sc and} {\sc
  Barr, A.~H.} 2003.
\newblock {\em Discrete differential-geometry operators for triangulated
  2-manifolds}.
\newblock Springer Berlin Heidelberg, Berlin, Heidelberg, 35--57.

\bibitem[\protect\citeauthoryear{Neumann, Varanasi, Theobalt, Magnor, and
  Wacker}{Neumann et~al\mbox{.}}{2014}]{neumann2014compressed}
{\sc Neumann, T.}, {\sc Varanasi, K.}, {\sc Theobalt, C.}, {\sc Magnor, M.},
  {\sc and} {\sc Wacker, M.} 2014.
\newblock Compressed manifold modes for mesh processing.
\newblock {\em Comput. Graph. Forum\/}~{\em 33,\/}~5 (Aug.), 35--44.

\bibitem[\protect\citeauthoryear{Ovsjanikov, Ben-Chen, Solomon, Butscher, and
  Guibas}{Ovsjanikov et~al\mbox{.}}{2012}]{ovsjanikov2012functional}
{\sc Ovsjanikov, M.}, {\sc Ben-Chen, M.}, {\sc Solomon, J.}, {\sc Butscher,
  A.}, {\sc and} {\sc Guibas, L.} 2012.
\newblock Functional maps: A flexible representation of maps between shapes.
\newblock {\em ACM Trans. Graph.\/}~{\em 31,\/}~4 (July), 30:1--30:11.

\bibitem[\protect\citeauthoryear{Ovsjanikov, Mérigot, Mémoli, and
  Guibas}{Ovsjanikov et~al\mbox{.}}{2010}]{ovsjanikov2010one}
{\sc Ovsjanikov, M.}, {\sc Mérigot, Q.}, {\sc Mémoli, F.}, {\sc and} {\sc
  Guibas, L.} 2010.
\newblock One point isometric matching with the heat kernel.
\newblock {\em Computer Graphics Forum\/}~{\em 29,\/}~5, 1555--1564.

\bibitem[\protect\citeauthoryear{Ozolins, Lai, Caflisch, and Osher}{Ozolins
  et~al\mbox{.}}{2013}]{Ozolins2013}
{\sc Ozolins, V.}, {\sc Lai, R.}, {\sc Caflisch, R.}, {\sc and} {\sc Osher, S.}
  2013.
\newblock Compressed modes for variational problems in mathematics and physics.
\newblock {\em Proceedings of the National Academy of Sciences\/}~{\em
  110,\/}~46, 18368--18373.

\bibitem[\protect\citeauthoryear{Pinkall and Polthier}{Pinkall and
  Polthier}{1993}]{pinkall1993computing}
{\sc Pinkall, U.} {\sc and} {\sc Polthier, K.} 1993.
\newblock Computing discrete minimal surfaces and their conjugates.
\newblock {\em Experiment. Math.\/}~{\em 2,\/}~1, 15--36.

\bibitem[\protect\citeauthoryear{Pokrass, Bronstein, Bronstein, Sprechmann, and
  Sapiro}{Pokrass et~al\mbox{.}}{2013}]{pokrass2013sparse}
{\sc Pokrass, J.}, {\sc Bronstein, A.~M.}, {\sc Bronstein, M.~M.}, {\sc
  Sprechmann, P.}, {\sc and} {\sc Sapiro, G.} 2013.
\newblock Sparse modeling of intrinsic correspondences.
\newblock {\em Computer Graphics Forum\/}~{\em 32,\/}~2pt4, 459--468.

\bibitem[\protect\citeauthoryear{Rustamov}{Rustamov}{2007}]{rustamov2007laplace}
{\sc Rustamov, R.~M.} 2007.
\newblock Laplace-beltrami eigenfunctions for deformation invariant shape
  representation.
\newblock In {\em Proceedings of the Fifth Eurographics Symposium on Geometry
  Processing}. SGP '07. Eurographics Association, Aire-la-Ville, Switzerland,
  Switzerland, 225--233.

\bibitem[\protect\citeauthoryear{Schwartz, Shaw, and Wolfson}{Schwartz
  et~al\mbox{.}}{1989}]{schwartz1989numerical}
{\sc Schwartz, E.~L.}, {\sc Shaw, A.}, {\sc and} {\sc Wolfson, E.} 1989.
\newblock A numerical solution to the generalized mapmaker's problem:
  Flattening nonconvex polyhedral surfaces.
\newblock {\em IEEE Trans. Pattern Anal. Mach. Intell.\/}~{\em 11,\/}~9
  (Sept.), 1005--1008.

\bibitem[\protect\citeauthoryear{Shtern and Kimmel}{Shtern and
  Kimmel}{2014a}]{shterniterative}
{\sc Shtern, A.} {\sc and} {\sc Kimmel, R.} 2014a.
\newblock Iterative closest spectral kernel maps.
\newblock In {\em 3D Vision (3DV), 2014 2nd International Conference on}.
  Vol.~1. IEEE, 499--505.

\bibitem[\protect\citeauthoryear{Shtern and Kimmel}{Shtern and
  Kimmel}{2014b}]{shtern2013matching}
{\sc Shtern, A.} {\sc and} {\sc Kimmel, R.} 2014b.
\newblock Matching the {LBO} eigenspace of non-rigid shapes via high order
  statistics.
\newblock {\em Axioms\/}~{\em 3,\/}~3, 300--319.

\bibitem[\protect\citeauthoryear{Shtern and Kimmel}{Shtern and
  Kimmel}{2015}]{shtern2015spectral}
{\sc Shtern, A.} {\sc and} {\sc Kimmel, R.} 2015.
\newblock Spectral gradient fields embedding for nonrigid shape matching.
\newblock {\em Computer Vision and Image Understanding\/}~{\em 140}, 21 -- 29.

\bibitem[\protect\citeauthoryear{{Simon}}{{Simon}}{2005}]{feyman_kack}
{\sc {Simon}, B.} 2005.
\newblock {\em {Functional integration and quantum physics. 2nd ed.}\/}, 2nd
  ed. ed.
\newblock Providence, RI: AMS Chelsea Publishing.

\bibitem[\protect\citeauthoryear{Sun, Ovsjanikov, and Guibas}{Sun
  et~al\mbox{.}}{2009}]{sun2009concise}
{\sc Sun, J.}, {\sc Ovsjanikov, M.}, {\sc and} {\sc Guibas, L.} 2009.
\newblock A concise and provably informative multi-scale signature based on
  heat diffusion.
\newblock In {\em Proceedings of the Symposium on Geometry Processing}. SGP
  '09. Eurographics Association, Aire-la-Ville, Switzerland, Switzerland,
  1383--1392.

\bibitem[\protect\citeauthoryear{Touma and Gotsman}{Touma and
  Gotsman}{1998}]{touma}
{\sc Touma, C.} {\sc and} {\sc Gotsman, C.} 1998.
\newblock Triangle mesh compression.
\newblock {\em PROC GRAPHICS INTERFACE. pp. 26-34. 1998\/}.

\bibitem[\protect\citeauthoryear{Vallet and Levy}{Vallet and
  Levy}{2008}]{vallet2008spectral}
{\sc Vallet, B.} {\sc and} {\sc Levy, B.} 2008.
\newblock {Spectral Geometry Processing with Manifold Harmonics}.
\newblock {\em Computer Graphics Forum\/}.

\bibitem[\protect\citeauthoryear{Van Der~Aa, Ter~Morsche, and Mattheij}{Van
  Der~Aa et~al\mbox{.}}{2007}]{van2007computation}
{\sc Van Der~Aa, N.}, {\sc Ter~Morsche, H.}, {\sc and} {\sc Mattheij, R.} 2007.
\newblock Computation of eigenvalue and eigenvector derivatives for a general
  complex-valued eigensystem.
\newblock {\em Electronic Journal of Linear Algebra\/}~{\em 16,\/}~1, 300--314.

\bibitem[\protect\citeauthoryear{Wei, Huang, Ceylan, Vouga, and Li}{Wei
  et~al\mbox{.}}{2016}]{haoli_CNN}
{\sc Wei, L.}, {\sc Huang, Q.}, {\sc Ceylan, D.}, {\sc Vouga, E.}, {\sc and}
  {\sc Li, H.} 2016.
\newblock Dense human body correspondences using convolutional networks.
\newblock In {\em Computer Vision and Pattern Recognition (CVPR)}.

\bibitem[\protect\citeauthoryear{Weyl et~al\mbox{.}}{Weyl
  et~al\mbox{.}}{1950}]{weyl1950ramifications}
{\sc Weyl, H.} {\sc et~al\mbox{.}} 1950.
\newblock Ramifications, old and new, of the eigenvalue problem.
\newblock {\em Bulletin of the American Mathematical Society\/}~{\em 56,\/}~2,
  115--139.

\bibitem[\protect\citeauthoryear{Zigelman, Kimmel, and Kiryati}{Zigelman
  et~al\mbox{.}}{2002}]{zigelman2002texture}
{\sc Zigelman, G.}, {\sc Kimmel, R.}, {\sc and} {\sc Kiryati, N.} 2002.
\newblock Texture mapping using surface flattening via multidimensional
  scaling.
\newblock {\em IEEE Transactions on Visualization and Computer Graphics\/}~{\em
  8,\/}~2 (Apr), 198--207.

\end{thebibliography}

\appendix
\section{Proof of Theorem 2.}

Let us be given the Hamiltonian operator 
 $H=-\Delta+V$.

Recall the Courant–Fischer min-max principle;
see also \cite{aflalo2014optimality} and
 \cite{brezis2010functional} Problems 37 and 49. We have for every $i \geq 0$, 
\begin{equation}\label{eq:proofOpt}
\begin{aligned}
E_{i+1}=\underset{\substack{
\Lambda\\
\text{codim} \Lambda = i}}{\text{max}}  \ \      
 \underset{\substack{
 f\in \Lambda\\
 f \neq 0}}{\text{min}} 
  \bigg \{ \frac{ \|\nabla f\|_{2}^{2}+\|\sqrt{V}f\|_{2}^{2}}{\|f\|_{2}^{2}}. 
    \bigg \}
\end{aligned}
\end{equation}
That is, the min is taken over a linear subspace $\Lambda \subset H1(S)$ with $H^{1}(S)$ is the Sobolev space
$\{f \in L^2, \nabla f \in L^2\}$ of co-dimension $i$ and the max is taken over all such subspaces.\\
Set $\Lambda_0 = \{f \in H^1(S); \langle f,\psi_k\rangle = 0, k = 1, 2,...,i\}$, so that $\Lambda_{0}$ is a subspace of co-dimension
$i$.\\ 
By \ref{eq:proofOpt} we have that for all $f \neq 0, f \in \Lambda_{0}$
\begin{equation}
\begin{aligned}
\frac{ \|\nabla f\|_{2}^{2}+\|\sqrt{V}f\|_{2}^{2}}{\|f\|_{2}^{2}}\geq \frac{E_{i+1}}{\alpha},
\end{aligned}
\end{equation}
and thus
\begin{equation}\label{eq:proofOpt2}
\begin{aligned}
X_{0}=\underset{\substack{
 f\in \Lambda\\
 f \neq 0}}{\text{min}}  \ \      
\frac{ \|\nabla f\|_{2}^{2}+\|\sqrt{V}f\|_{2}^{2}}{\|f\|_{2}^{2}}\geq \frac{E_{i+1}}{\alpha}.
\end{aligned}
\end{equation}
On the other hand, by \ref{eq:proofOpt},
\begin{equation}\label{eq:proofOpt3}
\begin{aligned}
E_{i+1}\geq X_{0}.
\end{aligned}
\end{equation}
Combining \ref{eq:proofOpt2} and \ref{eq:proofOpt3} yields $\alpha \geq 1$.       \ \ \ \ \ \ \  \ \ \ \ \ \ \ \ \ \ \ \ \ \ \ \ \ \ \ \ \ \ \ \ \ \ \ \ \ \ \ \ \ \ \ \ \ \ $\Box$

\section{Diffusion kernel of the Hamiltonian  }
In order to solve the diffusion equation, we first need to find the fundamental solution 
 kernel $K(x,y,t)$ to the Dirichlet problem that yields the heat
  equation
\begin{equation}\label{eq:heatKernelHDev1}
\Bigg\{ \begin{aligned}
&\partial_{t}K(x,y,t) =H(K(x,y,t))\cr
&\lim_{t\to 0}K(x,y,t) = \delta_{y}(x).
\end{aligned}
\end{equation} 
Recall that for $V=0$ we return to the regular LBO diffusion case.\\ Suppose that $H$  has a eigendecomposition
 $ \{ \psi_{i},E_{i} \} ^{ \infty}_{i=1}$. In that case, we can write 
\begin{equation}\label{eq:heatKernelHDev12}
\begin{aligned}
K(x,y,t) = \sum_{i}\langle K(x,y,t),\psi_{i}(x)\rangle_{\mathcal{M}} \psi_{i}(x)=\sum_{i}\alpha_{i}(t)\psi_{i}(x),
\end{aligned}
\end{equation}
 and from the linearity of $H$ we have
\begin{equation}\label{eq:heatKernelHDev2}
 \left\{ \begin{aligned}
& H(K(x,y,t))=\sum_{i}\alpha_{i}(t)H(\psi_{i})= \sum_{i}-E_{i}\alpha_{i}(t)\psi_{i}\\
&\partial_{t}K(x,y,t) = \sum_{i}\partial_{t}\alpha_{i}(t)\psi_{i}.
\end{aligned}\right .
\end{equation}
Since $\langle \psi_{i},\psi_{j}\rangle_{\mathcal{M}}=\delta_{ij} $, we have from (\ref{eq:heatKernelHDev1}) and (\ref{eq:heatKernelHDev2}) 
\begin{equation}\label{eq:heatKernelHDev3}
\begin{aligned}
 \partial_{t}\alpha_{i}(t) = -E_{i}\alpha_{i}(t),
\end{aligned}
\end{equation}
that leads to
\begin{equation}\label{eq:heatKernelHDev3_2}
\begin{aligned}
\alpha_{i}(t) = \alpha_{i}(0)e^{-E_{i}t}.
\end{aligned}
\end{equation}
As $\delta_{y}(x)=\sum_{i}\psi_{i}(y)\psi_{i}(x)$, 
 from the initial condition $K(x,y,0) =\delta_{y}(x)$, we obtain
\begin{equation}\label{eq:heatKernelHDev4}
\begin{aligned}
 &K(x,y,0) =\sum_{i}\alpha_{i}(0)\psi_{i}(x) = \sum_{i}\psi_{i}(y) \psi_{i}(x) =\delta_{y}(x)\\
&\Leftrightarrow \alpha_{i}(0)=\psi_{i}(y)\\
 &\Rightarrow K(x,y,t)=\sum_{i}e^{-E_{i}t}\psi_{i}(x)\psi_{i}(y).
\end{aligned}
\end{equation}
The solutions have the form 
\begin{equation}
\label{eq:heatKernelSol}
 u(x,t) = \int_{\mathcal{M}}u_{0}(y)K(x,y,t)da(y).
\end{equation}
\hfill\(\Box\)
 
\end{document}